\documentclass[aps,pre,superscriptaddress]{revtex4}

\usepackage{graphicx}
\usepackage{amssymb}
\usepackage{multirow}

\def\smallwhitecircle{\mbox{\large $\circ$}}
\def\smallblackcircle{\mbox{\large $\bullet$}}
\def\smallwhitesquare{\mbox{\scriptsize $\square$}}
\def\smallblacksquare{\mbox{\scriptsize $\blacksquare$}}

\def\smallwhitetriangleright{\mbox{\small $\vartriangleright$}}
\def\smallblacktriangleright{\mbox{\small $\blacktriangleright$}}

\def\smallstar{\mbox{\large$\ast$}}

\newcommand{\T}{\textstyle}

\def\3{\ss }           
\def\bfr{{\bf r}}
\def\bfk{{\bf k}}
\def\bfg{{\bf g}}

\def\bomega{\mbox{\boldmath $\omega$}}

\def\bxi{\mbox{\boldmath $\xi$}}
\def\bsxi{\mbox{\boldmath \tiny $\xi$}}
\def\rt{\bfr, t}

\def\be{\begin{equation}}
\def\ee{\end{equation}}
\def\bea{\begin{eqnarray}}
\def\eea{\end{eqnarray}}

\def\d{\delta}
\def\T{\Theta}
\def\z{\zeta}

\newcommand{\idelta}[1]{\int_{-a/2}^{a/2} d \delta_{#1}}
\newcommand{\ix}[1]{\int_{\delta_1}^{a+\delta_1} d x_{#1}}
\newcommand{\iX}[1]{\int_{0}^{a} d X_{#1}}

\begin{document}

\title{
Dynamic correlations in stochastic rotation dynamics} 
\author{E. T{\"u}zel}
\affiliation{School of Physics and Astronomy, 
University of Minnesota, \\ 116 Church Street SE, Minneapolis, MN 55455}
\affiliation{Supercomputing Institute, University of Minnesota, \\
599 Walter Library, 117 Pleasant Street S.E., \\
Minneapolis, MN 55455}
\author{T. Ihle}
\affiliation{Department of Physics, North Dakota State University, 
Fargo, ND 58105}
\author{D.M. Kroll}
\affiliation{Department of Physics, North Dakota State University,  
Fargo, ND 58105}

\begin{abstract}

The dynamic structure factor, vorticity and entropy density dynamic correlation 
functions are measured for Stochastic Rotation Dynamics (SRD), a particle based algorithm 
for fluctuating fluids. This allows us to obtain unbiased values for the 
longitudinal transport coefficients such as thermal diffusivity and bulk viscosity.
The results are in good agreement with earlier numerical and theoretical results, and 
it is shown for the first time that the bulk viscosity is indeed zero for this algorithm. 
In addition, corrections to the self-diffusion coefficient and shear viscosity 
arising from the breakdown of the molecular chaos approximation at small mean free 
paths are analyzed. In addition to deriving the form of the leading correlation 
corrections to these transport coefficients, the probabilities that two and three 
particles remain collision partners for consecutive time steps are derived 
analytically in the limit of small mean free path. The results of this paper verify 
that we have an excellent understanding of the SRD algorithm at the kinetic level 
and that analytic expressions for the transport coefficients derived elsewhere 
do indeed provide a very accurate description of the SRD fluid.
\end{abstract}

\maketitle
PACS number(s): 47.11.+j, 05.40.-a, 02.70.Ns 

\section{Introduction}

Several years ago, Malevanets and Kapral 
\cite{malevanets_99_mms,malevanets_00_smd} derived a simple and appealing 
algorithm---often called Stochastic Rotation Dynamics (SRD) or Multi-Particle 
Collision Dynamics (MPCD)---for the mesoscale modeling of fluctuating 
fluids. SRD is a particle based simulation technique with simple 
discrete time dynamics consisting of consecutive streaming and collision 
steps. It shares many features with Bird's Direct Simulation Monte Carlo 
(DSMC) algorithm~\cite{bird_94_mgd}, but uses more efficient multi-particle 
collisions to exchange momentum between the particles. Since there is a 
Boltzmann $H$-theorem for the SRD algorithm, and the  
particle number, momentum, and energy are locally conserved, 
the correct hydrodynamic behavior is guaranteed at large length and time 
scales. The algorithm therefore provides a convenient computational 
tool for solving the underlying thermo-hydrodynamic equations by  
providing a ``hydrodynamic heat bath'' which incorporates thermal 
fluctuations and provides the correct hydrodynamic interactions between 
embedded particles or polymers. An important advantage of SRD is that its 
simplified dynamics has enabled the analytical calculation of the transport 
coefficients and made it possible to obtain a rather complete theoretical 
understanding of the time-dependent correlation functions, the relaxation 
to equilibrium~\cite{ihle_01_srd,pooley_05_ktd}, and the behavior in shear flow, 
including shear thinning at high shear rates~\cite{ryder_05_thesis}.
Because the algorithm correctly includes long-ranged hydrodynamic 
interactions and Brownian fluctuations---both of which are generally 
required for a proper statistical treatment of the dynamics of mesoscopic 
suspended particles---it has been used to study the behavior of  
polymers~\cite{kikuchi_02_pcp,winkler_04_rcp,webster_05_mtp}, colloids~\cite{falck_04_ihm,inoue_02_dsm}, including 
sedimentation~\cite{padding_04_hbf, hecht_05_scc,padding_06_hib}, and 
vesicles in flow~\cite{noguchi_04_fvw,noguchi_05_dfv}.

In its original form \cite{malevanets_99_mms,malevanets_00_smd}, 
the SRD algorithm was not Galilean invariant at low temperatures, 
where the mean free path, $\lambda$, is smaller than the cell size $a$. 
However, it was shown \cite{ihle_01_srd,ihle_03_srd_a} 
that Galilean invariance can be restored by introducing a random shift 
of the computational grid before every collision.
In addition to restoring Galilean invariance, this grid shifting procedure
accelerates momentum transfer between cells and leads to a collisional
contribution to the transport coefficients. Two approaches
have been used to analyze the resulting algorithm and determine the shear
viscosity and thermal diffusivity. In Refs. \cite{pooley_05_ktd} 
and \cite{kikuchi_03_tcm}, a
non-equilibrium kinetic approach is used to derive the transport coefficients.
In Refs. \cite{ihle_01_srd,ihle_03_srd_a},
a discrete-time projection operator technique was utilized
to obtain Green-Kubo relations for the model's transport coefficients,
and explicit expressions for the transport coefficients were
derived in accompanying papers 
\cite{ihle_03_srd_b,tuzel_03_tcs,ihle_04_rgr,ihle_05_ect}.
The two approaches are complementary and, for the most part,
agree in their conclusions. The first is rather straightforward
and intuitively appealing, but makes several assumptions which are not easily
verified. The projection operator approach justifies in detail several
assumptions used in the non-equilibrium calculations of Refs. 
\cite{pooley_05_ktd} and \cite{kikuchi_03_tcm}; it can also be used to analyze
the transport coefficients of the
longitudinal modes, namely the bulk viscosity and thermal diffusivity, which
are hard to calculate analytically in the non-equilibrium approach
\cite{pooley_05_ktd}. Note, in particular, that the collisional contribution to
the thermal conductivity has not yet been determined using the non-equilibrium
methods.

In spite of some claims to the contrary~\cite{pooley_05_ktd}, both approaches yield 
the same results for the transport coefficients. Table \ref{tab_transport} 
contains a summary of both the collisional and kinetic contributions 
to the transport coefficients, including references to the original source 
of the results;  the table caption also provides a brief synapsis of 
some misprints in published articles which may lead to confusion.  
Simulation results have generally been in good agreement 
with these predictions. This is particularly true for the shear viscosity, 
where the results of equilibrium measurements of vorticity fluctuations 
\cite{ihle_03_srd_a,ihle_01_srd} and the Green-Kubo relations 
\cite{ihle_03_srd_b,tuzel_03_tcs} are in excellent agreement with 
non-equilibrium measurements in shear~\cite{kikuchi_03_tcm,pooley_05_ktd}
and Poiseuille ~\cite{allahyarov_02_mss} flow. The situation with the 
transport coefficients of the longitudinal modes, namely the bulk 
viscosity and thermal diffusivity, is somewhat less clear. The only 
reliable measurements of the thermal diffusivity 
have entailed equilibrium measurements of the corresponding Green-Kubo 
relation~\cite{ihle_03_srd_b,tuzel_03_tcs,ihle_04_rgr,ihle_05_ect} and a 
non-equilibrium measurement obtained by setting up a temperature gradient 
and measuring the resulting energy flux in a regime where the 
collisional contribution is negligible~\cite{pooley_05_ktd}. To our 
knowledge, there has been no direct verification of the prediction that 
the bulk viscosity is zero for SRD. 
   
In this paper we take an alternative approach based on an analysis of the 
equilibrium fluctuations of the hydrodynamic modes to directly measure 
the shear and bulk viscosities and thermal diffusivity. 
Starting with an analysis of vorticity fluctuations to determine 
the shear viscosity, measurements of the dynamic structure factor 
are then used to deduce the values of the speed of sound, the thermal 
diffusivity, and the bulk 
viscosity. Measurements of the temporal behavior of the entropy 
correlations are also used to obtain a direct independent measurement 
of the thermal diffusivity. To our knowledge, this is 
the first quantitative measurement of the dynamic structure factor for SRD. 
An earlier measurement by Inoue et al. \cite{inoue_03_dcs} lead to unphysical 
results in the large frequency limit, and could not be used to determine 
the transport coefficients.

The results of these measurements verify directly that the bulk viscosity 
is indeed zero for this algorithm. In addition, in agreement with earlier 
work \cite{ihle_04_rgr,tuzel_03_tcs,ihle_03_srd_a,ihle_03_srd_b,pooley_05_ktd,kikuchi_03_tcm}, 
results for the shear viscosity and the thermal diffusivity are in excellent 
agreement with the theoretical predictions presented in Table 
\ref{tab_transport} for a wide range 
of particle densities and mean free paths. However, as noted originally 
in Ref. \cite{ihle_03_srd_b}, and discussed in more detail in Ref. 
\cite{ihle_04_rgr}, correlations between particles occupying the same 
collision cell at different time steps lead to an enhanced kinetic 
contribution to the transport coefficients. This breakdown of the molecular 
chaos approximation becomes pronounced at small mean free path, $\lambda$,  
since particles do not travel far between collisions and 
tend to repeatedly have the same collision partners. For most transport 
coefficients, this additional contribution to the transport coefficients 
is masked by the collisional contribution, which dominates in the small 
mean free path regime. The affect is particularly pronounced, however, 
for the self-diffusion coefficient, for which there is no collisional 
contribution. Indeed, Ripoll et al. \cite{ripoll_04_lhc,ripoll_05_drf} have 
observed that the self diffusion coefficient is significantly larger than 
the theoretical prediction of Ref. \cite{tuzel_03_tcs,ihle_03_srd_a} for small 
$\lambda/a$. Ripoll et al. provided a semi-analytical description of this 
behavior in which they determined numerically the number of particles sharing
the same cell as a function of time.
In this paper, we provide a detailed discussion of the leading correlation 
corrections to the kinetic contribution of both the shear viscosity 
and the self-diffusion coefficient and determine analytically the probability 
that two and three particles are in the same collision cells for consecutive 
time steps. While our results for the self-diffusion coefficient are in general 
agreement with those of Ref. \cite{ripoll_05_drf}, there seem to be 
several misprints in \cite{ripoll_05_drf}, making a detailed comparison 
difficult.

The remainder of the paper is organized as follows. After a brief summary 
of the SRD algorithm in Sec. II, the hydrodynamic equations of a simple 
liquid are reviewed and the correct form of the constitutive equations 
are discussed in Sec. III. The consequences of the fact that angular 
momentum is not conserved in the SRD algorithm are summarized, and the 
correct form of the viscous stress tensor is discussed. In particular, 
it is emphasized that in two dimensions, there is no difference between 
the viscous stress tensor of a simple isotropic fluid and an SRD 
fluid. The slight differences in three dimensions leave the form of 
the Navier-Stokes unchanged, with only a reinterpretation of the coefficient 
of sound attenuation. Sec. IV contains a fairly detailed derivation of the  
dynamic correlation functions in a simple liquid. The discussion follows 
rather closely that of Ref. \cite{forster_75_hfb}, but is included because 
several aspects of the derivation are a bit subtle and are generally not 
addressed in the literature. Explicit expressions for the vorticity, density, 
and entropy density dynamic correlations functions are presented. In 
Sec. V, these results are used to determine the shear and bulk viscosities  
and the thermal diffusivity. The agreement with the theoretical predictions 
summarized in Table \ref{tab_transport} is excellent. Sec. VI contains 
a detailed discussion of the consequence of the breakdown of the 
molecular chaos approximation at short mean free paths. While correlation 
effects do not change the collisional contributions to the transport 
coefficients~\cite{ihle_04_rgr}, they do dramatically increase the amplitude of the kinetic 
contribution. In addition to deriving the form of the leading correlation 
contributions to the shear viscosity and the self-diffusion coefficient, the 
probabilities that two ($p_2$) and three ($p_3$) particles are in the 
same collision cells at consecutive time steps are derived analytically in 
the limit $\lambda/a\to0$. More generally, simulation results are used to 
show that $p_2$ is solely a function of $\lambda/a$. In the case of the 
shear viscosity, it is shown that inclusion of the leading correlation 
corrections yields results in surprisingly good agreement with measurements 
of the viscous stress correlations. For the self-diffusion 
coefficient, however, 
the correlation corrections for larger time intervals are large, and 
dramatically increase the measured value of the self-diffusion coefficient 
for mean free paths smaller than the cell size. It is important to 
note that the correlation corrections considered here---arising from 
particles which collide with the same particles in consecutive time 
steps---are similar to those which occur in dense fluids interacting through 
soft potentials, and should therefore be interpreted as a ``potential'' or 
``collisional'' contribution to the velocity or stress 
correlation functions rather than a precursor of the power-law (long-time) 
tails observable at longer times. We believe that it is important 
to distinguish 
between these two effects, since long time tails are also visible at large 
mean free paths where these corrections are negligibly small. 
Although the same approach can be used to calculate these contributions, 
the corresponding probabilities are much harder to estimate. 
 
The results of this paper verify that we have an excellent 
understanding of the SRD algorithm at the kinetic level and that---with 
the exception of the self-diffusion coefficient---the analytic expressions 
for the transport coefficients given in Table \ref{tab_transport} do indeed 
provide a very accurate description of the SRD fluid. Furthermore, 
the analysis of the dynamical structure factor and the dynamic entropy density 
correlation function verify directly that the algorithm satisfies the 
fluctuation-dissipation theorem. While this is to be expected for the current 
algorithm, which satisfies the necessary semi-detailed balance conditions~\cite{malevanets_99_mms,
ihle_03_srd_a}, 
verification studies of this type will be important for generalizations 
of the algorithm which model excluded volume effects through the use of 
biased multi-particle collision rules which depend on the local velocities 
and densities~\cite{ihle_06_cpa,tuzel_06_ctc}. 

\section{Model}

In the SRD algorithm, the fluid is modeled by particles with continuous 
spatial coordinates ${\bf r}_i(t)$ and velocities ${\bf v}_i(t)$. 
The system is coarse-grained into the cells of a regular lattice with
no restriction on the number of particles in a cell. The evolution of
the system consists of two steps: streaming and collision. In the
streaming step, the coordinate of each particle is incremented by its
displacement during the time step, $\tau$. Collisions are
modeled by a simultaneous stochastic rotation of the relative velocities
of {\em every} particle in each cell. As discussed in Refs. \cite{ihle_01_srd} 
and \cite{ihle_03_srd_a}, a random shift of the particle coordinates 
(or, equivalently, the cell grid) before 
the collision step is required to ensure Galilean invariance. 
All particles are shifted by the {\em same} random vector with 
components in the interval $[-a/2,a/2]$ before the collision step. Particles 
are then shifted back to their original positions after the collision. If 
we denote the cell coordinate of the shifted particle $i$ by $\bxi_i^s$, 
the algorithm is summarized in the equations  
\bea
\label{eqm_1}
{\bf r}_i(t+\tau)&=&{\bf r}_i(t)+\tau\;{\bf v}_i(t) \\
\label{eqm_2}
{\bf v}_i(t+\tau)&=&{\bf u}[\bxi_i^s(t+\tau)]+\bomega[\bxi_i^s(t+\tau)]
\cdot\{{\bf v}_i(t)-{\bf u}[\bxi_i^s(t+\tau)]\},
\eea
where $\bomega(\bxi_i^s)$ denotes a stochastic rotation matrix, and 
${\bf u}(\bxi_i^s)\equiv{1\over M}\sum_{k\in\bsxi^s}{\bf v}_k$ is the 
mean velocity of the particles in cell $\bxi^s$. All particles in the 
cell are subject to the same rotation, but the rotations in different cells 
are statistically independent. There is a great deal of freedom in how the 
rotation step is implemented, and any stochastic rotation matrix consistent 
with detailed balance can be used. 
In two dimensions, the stochastic rotation matrix, $\bomega$, is typically 
taken to be a rotation by an angle $\pm\alpha$, with probability $1/2$ 
(see Refs. \cite{ihle_01_srd,ihle_03_srd_a,ihle_03_srd_b}). In three 
dimensions, one can perform rotations by an angle $\alpha$ about a randomly 
chosen direction, where all orientations of the random axis occur with 
equal probability (Model A in Ref.~\cite{tuzel_03_tcs}).
	
\section{Hydrodynamics and Transport coefficients}

There is a hydrodynamic mode associated with each conserved density in a 
fluid. For  a simple liquid, the conserved quantities are the particle mass 
density, $\hat\rho(\rt)$, the momentum density, $\hat{\bf g}(\rt)$, and the 
energy density, $\hat\epsilon (\rt)$, and the corresponding microscopic 
conservation laws 
are  
\be   \label{density}   
\partial_t\hat\rho(\rt) + \partial_\beta \hat g_\beta(\rt)=0,  
\ee 
\be \label{momentum}   
\partial_t\hat g_\alpha(\rt) + \partial_\beta\hat\tau_{\alpha\beta}(\rt)=0,   
\ee 
and 
\be \label{energy} 
\partial_t\hat\epsilon(\rt) + \partial_\beta \hat\chi_\beta(\rt)=0,   
\ee   
where $\hat\tau_{\alpha\beta}(\rt)$ are the Cartesian components of the 
microscopic stress tensor and $\hat{\chi}_\alpha(\rt)$ is the 
$\alpha$-component of the microscopic energy current density. 
While Eqs. (\ref{density})-(\ref{energy}) are 
microscopically exact, macroscopic constitutive relations are required  
to close the system of equations. The constitutive equations relate 
local non-equilibrium averages of $\hat{\bf g}(\rt)$, 
$\hat\tau_{\alpha\beta}(\rt)$, and $\hat{\chi}_\alpha(\rt)$ 
to the local hydrodynamic variables, 
$\rho(\rt)\equiv\langle\hat\rho(\rt)\rangle$, 
${\bf g}(\rt)\equiv\langle\hat{\bf g}(\rt)\rangle$, 
$\epsilon(\rt)\equiv\langle\hat\epsilon (\rt)\rangle$, and their gradients. 
For a simple isotropic liquid, the constitutive relations have the 
form \cite{forster_75_hfb, boon_92_mh}  
\bea     
\langle \hat g_\alpha (\rt) \rangle &=& g_\alpha(\rt )  ,  \label{ce1} \\
\langle \hat\tau_{\alpha\beta}(\rt) \rangle &=&  p (\rt) \delta_{\alpha\beta} - 
\sigma_{\alpha\beta}(\rt) \\
&=&p (\rt) \delta_{\alpha\beta} - 
\nu [\partial_\alpha g_\beta(\rt) + \partial_\beta g_\alpha(\rt) 
- (2/d)\delta_{\alpha\beta}\,\partial_\lambda g_\lambda\,] - 
\gamma \delta_{\alpha\beta}\,\partial_\lambda g_\lambda\, ,  \label{ce2}  \\
\langle \hat\chi_\alpha(\rt)\rangle& = &[(\epsilon + p)/\rho] g_\alpha(\rt) - 
\kappa \partial_\alpha T(\rt)\;\;,      \label{ce3}
\eea
where $p(\rt)$ and $T(\rt)$ are the local pressure and temperature fields, 
respectively. $p$ is the equilibrium pressure and $\epsilon$ 
is the equilibrium energy density; $\nu$ and $\gamma$ are the kinematic shear 
and bulk viscosities, respectively, and $\kappa$ is the thermal conductivity. 
$d$ is the spatial dimension, and $\sigma_{\alpha\beta}$ is the macroscopic 
viscous stress tensor. There are both non-derivative, reactive, and 
dissipative contributions to the constitutive relations. 
The form of the reactive 
terms can be inferred from Galilean invariance. The dissipative terms follow 
from an expansion of the current densities to first order in the gradients 
of the local conjugate forces ${\bf g}(\rt)$,  $p(\rt)$, and $T(\rt)$; 
symmetry dictates the general form of these terms.   
Non-linear terms in the constitutive equations 
have been omitted because we only require the linear hydrodynamic 
equations in the following. 

The local equilibrium averages of Eqs. (\ref{density})-(\ref{energy}), 
together with the constitutive relations given by Eqs. (\ref{ce1})-(\ref{ce3})  provide 
a complete description of the hydrodynamics of the liquid. 
The resulting linearized Navier-Stokes equation is 
\be\label{NS}   
\partial_t g_\alpha(\rt) + \partial_\alpha p(\rt) - 
\nu \partial_\lambda^2 g_\alpha (\rt)
- \left(\gamma + \frac{d-2}{d}\nu\right) 
\partial_\alpha \partial_\lambda g_\lambda(\rt) = 0 \;\;.    
\ee 
The corresponding equations for the mass density and energy density are  
\begin{equation}\label{dens_hydro} 
\partial_t\rho(\rt) + \partial_\beta g_\beta(\rt)=0,  
\end{equation}
and
\begin{equation}\label{energy_hydro}  
\partial_t\epsilon(\rt) + [(\epsilon+p)/\rho]\partial_\lambda g_\lambda(\rt) - 
\kappa\partial_\lambda^2 T(\rt) = 0,  
\end{equation}
respectively. 

Because of the cell structure introduced in SRD to define the collision 
environment, angular momentum is not conserved in a SRD collision. 
As a consequence, 
the macroscopic viscous stress tensor is not a symmetric function of the 
derivatives $\partial_\alpha g_\beta$, and instead of Eq. (\ref{ce2}), the 
constitutive equation has the general form~\cite{ihle_05_ect} 
\bea   
\label{SRD_ST} 
 \tau_{\alpha\beta}(\rt)  = p (\rt) \delta_{\alpha\beta} &-& 
\nu_1 \left[\partial_\alpha g_\beta(\rt) + \partial_\beta g_\alpha(\rt) 
- (2/d)\delta_{\alpha\beta} \partial_\lambda g_\lambda \right] \\ & - &
\nu_2(\partial_\beta g_\alpha(\rt) - \partial_\alpha g_\beta(\rt)) - 
\gamma \delta_{\alpha\beta} \partial_\lambda g_\lambda , 
\eea   
where $\nu_2$ is a new viscous transport coefficient associated with the 
non-symmetric part of the stress tensor. 
Because the kinetic contribution to the microscopic stress tensor is 
symmetric, $\nu_2^{kin}\equiv0$ and  $\nu_1^{kin}\equiv\nu^{kin}$. It is also 
easy to show that $\gamma^{kin} =0$, so that the kinetic contribution to 
the macroscopic viscous stress tensor is 
\be 
\sigma^{kin}_{\alpha\beta}(\rt)  = \nu^{kin}[\partial_\alpha g_\beta (\rt) 
+ \partial_\beta g_\alpha(\rt) - 
(2/d)\delta_{\alpha\beta}\partial_\lambda g_\lambda(\rt)]. 
\ee  
In Ref. \cite{ihle_05_ect}, it was also shown that the collisional 
contributions to the viscous transport coefficients fulfill the relation 
\be 
[(d-2)/d]\nu_1^{col} - \nu_2^{col} + \gamma^{col}=0,   
\ee 
and that the collision contribution to the macroscopic viscous stress tensor is 
\be 
 \sigma^{col}_{\alpha\beta}(\rt)  =(\nu_1^{col} + 
\nu_2^{col})\partial_\beta g_\alpha(\rt) = 
\nu^{col}\partial_\beta g_\alpha(\rt)\;\;, 
\ee 
up to a tensor with vanishing divergence, which will therefore not appear 
in the linearized hydrodynamic equations.    
The resulting linearized hydrodynamic equation for the momentum density is, 
therefore  
\begin{equation}\label{mom_eq}  
\partial_t g_\alpha(\rt) + \partial_\alpha p(\rt) - 
\nu\partial_\lambda^2 g_\alpha(\rt) - 
\frac{d-2}{d}\nu^{kin}\partial_\alpha \partial_\lambda g_\lambda(\rt) 
= 0, 
\end{equation}
where $\nu= \nu^{kin} + \nu^{col}$ and $\nu^{kin}$ and $\nu^{col}$ are the 
kinetic and collision contributions 
to the shear viscosity. The equations for the mass and energy densities 
remain unchanged. Comparison of Eq. (\ref{mom_eq}) with Eq. (\ref{NS}) 
shows that the only 
difference between the Navier-Stokes equation for an isotropic liquid and an 
SRD fluid is in the coefficient of the $\partial_\alpha \partial_\lambda 
g_\lambda(\rt)$ term, where $\nu$ is replaced by $\nu^{kin}$. The 
bulk viscosity does not appear in Eq. (\ref{mom_eq}) because it is zero for the 
SRD algorithm. Note that both equations are identical in $d=2$ and that   
the only difference in $d=3$ is a correction to the sound attenuation 
coefficient associated with the viscous dissipation of longitudinal sound 
waves. 



\section{Dynamic Correlations}

Spontaneous thermal fluctuations of the density, $\rho(\rt)$, momentum 
density, ${\bf g}(\rt)$, the energy density, $\epsilon (\rt)$ are dynamically 
coupled, and an analysis of their dynamic correlation functions in the 
limit of small wave vectors and frequencies 
can be used to determine a fluid's transport coefficients. In particular, 
because it is easily measured in dynamic light scattering, x-ray, and 
neutron scattering experiments, the density-density correlation 
function---the dynamic 
structure factor---is one of the most widely used vehicles for probing the 
dynamic and transport properties of liquids~\cite{berne_00_dls}.
 
In the following, we summarize the predictions of linearized hydrodynamics 
for the dynamic correlation functions 
of simple liquids in the hydrodynamic regime, and then use the results to 
analyze SRD simulation data in order to validate the theoretical results 
for the transport coefficients given in Table I. In particular, we provide 
in this way the first direct confirmation that the bulk viscosity is indeed 
zero for this model. 
Our discussion follows closely that of Ref. \cite{forster_75_hfb}, but is 
included because the derivation using Laplace transforms is not widely used 
in the literature, and the detailed results are needed in the subsequent 
analysis. The starting point is the linearized hydrodynamic equations 
given in Eqs. (\ref{NS}), (\ref{dens_hydro}), and (\ref{energy_hydro}). 
There are four modes, one transverse shear mode, and three coupled 
longitudinal modes. In order to keep the analysis general, we include the 
bulk viscosity in this section.\break 

\noindent {\it Transverse fluctuations}: Divide the momentum density 
${\bfg (\rt)}$ into transverse and longitudinal components,  
\be
\bfg(\rt)=\bfg_{\Vert}(\rt)+\bfg_{\perp}(\rt), 
\ee
where $\nabla \times \bfg_{\Vert}(\rt) = 0$ and 
$\nabla \cdot \bfg_{\perp}(\rt) =0$. Taking the curl of the Navier-Stokes 
equation, the transverse component of the momentum density, 
${\bf g}_\perp(\rt)$, 
is found to satisfy the diffusion equation
\be 
\partial_t\bfg_\perp(\rt) = \nu\partial_\lambda^2\bfg_\perp(\rt).
\ee  
By performing the Fourier-Laplace transform (Im~$z>0$) 
\be\label{gperpr}  
\bfg_{\perp}(\bfk,z)=\int_0^\infty dt e^{izt}
\int_V d\bfr\ e^{-i\bfk \cdot \bfr} \bfg_{\perp}(\rt)\;,  
\ee
the solution of the initial value problem, which describes the response of 
the transverse mode to an initial perturbation $\delta \bfg_\perp (\bfk, t=0)$ 
from equilibrium, is  
\be\label{omega_scalar}
\Omega_{{\bf g}_\perp}\delta \bfg_{\perp} (\bfk,z) =
i~ \delta \bfg_\perp (\bfk, t=0),   
\ee
where $\Omega_{{\bf g}_\perp}\equiv z+ik^2\nu$.\hfill\break   

\noindent {\it Longitudinal fluctuations}:
For the longitudinal components, it is convenient to introduce the variable $q(\rt)$, which is ($T$ times)
the entropy density,
\be
q(\rt)\equiv\epsilon (\rt) - \frac{\epsilon+p}{\rho}\rho(\rt)
\ee
in place of the energy density and use the relations 
\be 
\nabla p(\rt) = \left(\frac{\partial p}{\partial \rho}\right)_S
\nabla\rho(\rt) + \frac{V}{T}\left(\frac{\partial p}{\partial S}
\right)_\rho \nabla q(\rt) 
\ee 
and 
\be 
\nabla T(\rt) = \left(\frac{\partial T}{\partial \rho}\right)_S 
\nabla\rho(\rt) + \frac{V}{T}\left(\frac{\partial T}{\partial S}
\right)_\rho \nabla q(\rt), 
\ee 
where $S$ is the total entropy, to eliminate the pressure and temperature 
fields. Taking the Fourier-Laplace transform, the resulting coupled set 
of equations for the longitudinal modes can be written as  
\be \underbrace{\left( 
\begin{array}{ccc}
z     & -k           & 0                                                    \\
-k c^2 & z+ik^2D_\ell & -\frac{V}{T}\left(\frac{\partial p}{\partial S} \right)_\rho k \\
ik^2\kappa \left(\frac{\partial T}{\partial \rho } \right)_S  & 0  & 
z+ik^2\frac{\kappa}{\rho c_v} \end{array} \right) }_{\bf \Omega}
\left(\begin{array}{c}
\delta\rho(\bfk,z)   \\ \delta g_\Vert (\bfk,z)   \\  \delta q(\bfk,z)  \end{array} \right)= i 
\left(\begin{array}{c}
\delta\rho(\bfk,t=0)  \\  \delta g_\Vert (\bfk,t=0)   \\  \delta q(\bfk,t=0) \end{array} 
\right),  
\label{lmodes} 
\ee
where, for example, $\delta\rho(\bfk,t=0)\equiv\rho(\bfk,t=0)-\rho$; 
$\rho$ is the equilibrium density, and $k=\vert{\bf k}\vert$. Note that since 
$\delta \bfg_\Vert (\bfk,z) \parallel \bfk$, the equation for the longitudinal 
component of the momentum density is a scalar equation.
When writing Eq. (\ref{lmodes}), we have used the relations 
\be 
\rho c_v = \frac{T}{V}\left(\frac{\partial S}{\partial T}\right)_\rho,\ \ \ 
\ \ \ \ \rho c_p = \frac{T}{V}\left(\frac{\partial S}{\partial T}
\right)_p,\ \ \ \ {\rm and}\ \ \ 
c^2 = \left(\frac{\partial p}{\partial \rho}\right)_S = 
\frac{c_p}{c_v}\left(\frac{\partial p}{\partial \rho}\right)_T  
\ee 
to simplify the final expression. 
$c$ is the adiabatic speed of sound, and $D_\ell = 2[(d-1)/d]\nu + \gamma$. 
Eq. (\ref{lmodes}) describes how the longitudinal modes relax in response 
to initial perturbations $\delta\rho(\bfk,t=0)$, $\delta g_\Vert (\bfk,t=0)$, 
and $\delta q(\bfk,t=0)$. The zeros of the determinant of the coefficient 
matrix, $\bf \Omega$, give the complex frequencies of the hydrodynamic modes of the system. 
For small wave vector $\bfk$, the solutions of the resulting cubic equation are 
(up to terms of order $k^3$)  
\be 
\label{mode_s}
z = \pm ck - \frac{i}{2}k^2\Gamma\ \ \ \ \ \ \ {\rm (sound\ poles)} 
\ee 
and 
\be 
\label{mode_h}
z = -ik^2 D_T\ \ \ \ \ \ {\rm (heat\ pole)}, 
\ee 
where $D_T = \kappa/(\rho c_p)$ is the thermal diffusivity;  
$\Gamma=D_T (c_p/c_v-1) + D_\ell$, is the sound attenuation 
coefficient. In deriving Eqs. (\ref{mode_s}) and 
(\ref{mode_h}), the thermodynamic relation 
\be 
D_T\, c^2(c_p/c_v-1) = \kappa\,\frac{V}{T}\left(\frac{\partial T}{\partial\rho}
\right)_S\left(\frac{\partial p}{\partial S}\right)_\rho  
\ee 
has been used. 
\par

\noindent {\it Correlation functions}:
The matrix of dynamic correlation functions 
\be 
S_{ij}(\bfk,z) = \int_0^\infty d(t-t') \int_V d({\bf r}-{\bf r'}) 
e^{iz(t-t')-i\bfk\cdot({\bf r}-{\bf r'})} 
\langle [A_i(\bfr,t)-\langle A_i\rangle][A_j({\bf r'},t')-\langle A_j\rangle]  
\rangle 
\ee 
is given by \cite{forster_75_hfb} 
\be\label{lcf}  
S_{ij}(\bfk,z) = ik_BT({\bf \Omega}^{-1})_{il}\chi_{lj}(\bfk) , 
\ee 
where there is a sum over repeated indices. For the transverse modes, 
$\Omega$ is simply a scalar function defined in Eq. (\ref{omega_scalar}); 
for the longitudinal modes, however, it is a matrix, and the subscript indices 
denote the modes $\delta\rho(\bfk,t=0)$, $\delta g_\Vert (\bfk,t=0)$, 
and $\delta q(\bfk,t=0)$. $\chi_{lj}(\bfk)$ is the static susceptibility 
matrix. 

The correlation function for the transverse mode follows from 
Eqs. (\ref{omega_scalar}) and (\ref{lcf}) and 
$\chi_{g_\perp g_\perp} (\bfk)= \rho$, 
and is given by 
\be\label{cfgp}  
S_{g_\perp g_\perp}(k,z) = \frac{ik_BT\rho}{z+ik^2\nu}\;.   
\ee 
Taking the inverse Laplace transform, 
\be
S_{g_\perp g_\perp}(k,t) = \rho k_B T e^{-\nu k^2 t}\;\;.
\ee 
For the longitudinal modes, inverting ${\bf \Omega}$ and 
using the results  
$\chi_{\bfg_\Vert\rho} (\bfk)= 0$, $\chi_{\bfg_\Vert\bfg_\Vert} (\bfk)= \rho$,  
$\lim_{k\to0}\chi_{\rho\rho}(\bfk) = 
\rho(\partial\rho/\partial p)_T$, 
$\lim_{k\to0}\chi_{qq}(\bfk) = \rho T c_p$, and  
$\lim_{k\to0}\chi_{\rho q}(\bfk) = (T/m) (\partial\rho/\partial T)_p$  
(where $m$ is the particle mass), 
one finds 
\be  
S_{\rho\rho}(k,z)=ik_BT\rho\left(\frac{\partial \rho}{\partial p}\right)_T
\left[\frac{c_v}{c_p}
\left\{\frac{z+ik^2\left(\Gamma+D_T[c_p/c_v-1]\right)}{z^2-c^2k^2+izk^2\Gamma}\right\}
+\left(1-\frac{c_v}{c_p}\right)\frac{1}{z+ik^2D_T}\right] 
\label{crhorho}
\ee 
and 
\be 
S_{qq}(k,z)=ik_BT\frac{\rho c_pT}{z+ik^2D_T}\label{cqq}
\ee  
for the two scalar modes. 
Note that all non-vanishing static susceptibilities 
are symmetric in $k$, so that $k$-dependent corrections to $\chi_{ij}$ are 
$O(k^2)$ and therefore negligible in these expressions. 
For details, the reader is referred to Refs.~\cite{forster_75_hfb,boon_92_mh}.

The complete spectral transform of the time dependent density correlation 
function, the dynamic structure factor, $S_{\rho\rho}(k,\omega)$, 
is obtained by setting 
$z=\omega+i\delta$ and taking the limit $\delta\rightarrow 0$ 
in Eq. (\ref{crhorho}). The final result is \cite{forster_75_hfb},   
\be
S_{\rho\rho}(k,\omega) = 2 k_BT\rho\left(\frac{\partial \rho}
{\partial p}\right)_T \left[\frac{(c_v/c_p) c^2 k^4 \Gamma}
{\left(\omega^2-c^2k^2\right)^2+\left(\omega k^2 \Gamma\right)^2}+
\frac{(1-c_v/c_p) k^2 D_T}{\omega^2+\left(k^2D_T\right)^2}
-\left(1-\frac{c_v}{c_p}\right)\frac{\left(\omega^2-c^2k^2\right) k^2 D_T}
{\left(\omega^2-c^2k^2\right)^2+\left(\omega k^2 \Gamma\right)^2}\right]\;.   
\label{skw}
\ee

In experiments, it is genenerally not possible to measure 
$S_{\rho \rho}(k,t)$ directly. In simulations, however,
both $S_{\rho \rho}(k,t)$ and $S_{\rho\rho} (k,\omega)$ can be measured 
for a range of mean free paths and collision angles. 
The simplest way to determine $S_{\rho \rho}(k,t)$ is to take the 
inverse Laplace transform of Eq. (\ref{crhorho}). In light of Eqs. 
(\ref{mode_s}) and (\ref{mode_h}), it is sufficient to keep terms $O(k)$ in 
real parts and $O(k^2)$ in the imaginary parts when evaluating the resulting 
contour integral. The final result is 
 \be
S_{\rho \rho}(k,t)= 2 k_BT \rho\left(\frac{\partial \rho}
{\partial p}\right)_T  \left( \frac{c_v}{c_p}e^{-{\Gamma k^2 t/2} }
\hspace{-0.15cm}\left[ \cos\left(c k t\right)+
\left(\frac{\Gamma}{2}+\left(\frac{c_p}{c_v}-1\right) D_T\right) \frac{k}{c}  
\sin\left(c k t\right) \right]  
+\left(1-\frac{c_v}{c_p}\right) e^{-D_T k^2 t} \right). \label{skt}
\ee

Using Eq. (\ref{cqq}), it can be shown that the correlation function for  
the entropy density, 
$q(\rt)$, is given by
\be
S_{qq}(k,t)=\rho c_p k_BT^2 e^{-D_T k^2 t}. \label{sqq}
\ee
Note that Eq. (\ref{sqq}) provides an independent way to directly measure $D_T$. 

These results remain valid for the SRD fluid. The only modification is that 
$D_\ell = 2\left[(d-1)/d\right]\nu^{kin}+\nu^{col}$, so that the sound 
attenuation coefficient is 
\begin{equation}
\Gamma=D_T \left(\frac{c_p}{c_v}-1\right)+2 \left(\frac{d-1}{d} \right)
\nu^{kin}+\nu^{col}. 
\end{equation}
Note that in two-dimensions, the sound attenuation coefficient  
for a SRD fluid has the same functional dependence on $D_T$ and 
$\nu\equiv\nu^{kin}+\nu^{col}$ as an isotropic fluid with an ideal gas 
equation of state (for which $\gamma=0$). Finally, since SRD describes an 
ideal fluid, $p = \rho k_BT/m$ and $c_p = k_B/m + c_v = (d+2)k_B/2m$.  

\section{Measurements}

In our SRD simulations in two dimensions, the mass, momentum, and energy 
densities are measured at the cell level. The cell densities, $A^c(\bxi,t)$, 
are defined at the discrete set of coordinates $\bxi=a {\bf m}$, with 
$m_\beta=1,\ldots,L$, for each spatial dimension \cite{ihle_03_srd_a}. 
A superscript $c$ will be used to denote that the corresponding 
quantity is defined at the cell level. 
The Fourier transform of the cell variables are 
\be
A^c(\bfk,t)=\sum_{\bsxi} A^c(\bxi,t) e^{i \bfk \cdot \bsxi},   
\ee
and the inverse transform is 
\be
A^c(\bxi,t)=\frac{1}{L^d}\sum_{\bfk} A^c(\bfk,t)e^{-i \bfk \cdot \bsxi}. 
\ee
The Fourier-Laplace transforms of the corresponding dynamic 
correlation functions are  
\be 
S^c_{ij}(\bfk,z) = \int_0^\infty d(t-t') \sum_{\bsxi} 
e^{iz(t-t')-i\bfk \cdot \bsxi} 
\langle [A^c_i(\bxi,t)-\langle A^c_i\rangle]
[A^c_j({\bf 0},t')-\langle A^c_j\rangle]
\rangle \;\;.
\ee

\begin{figure}
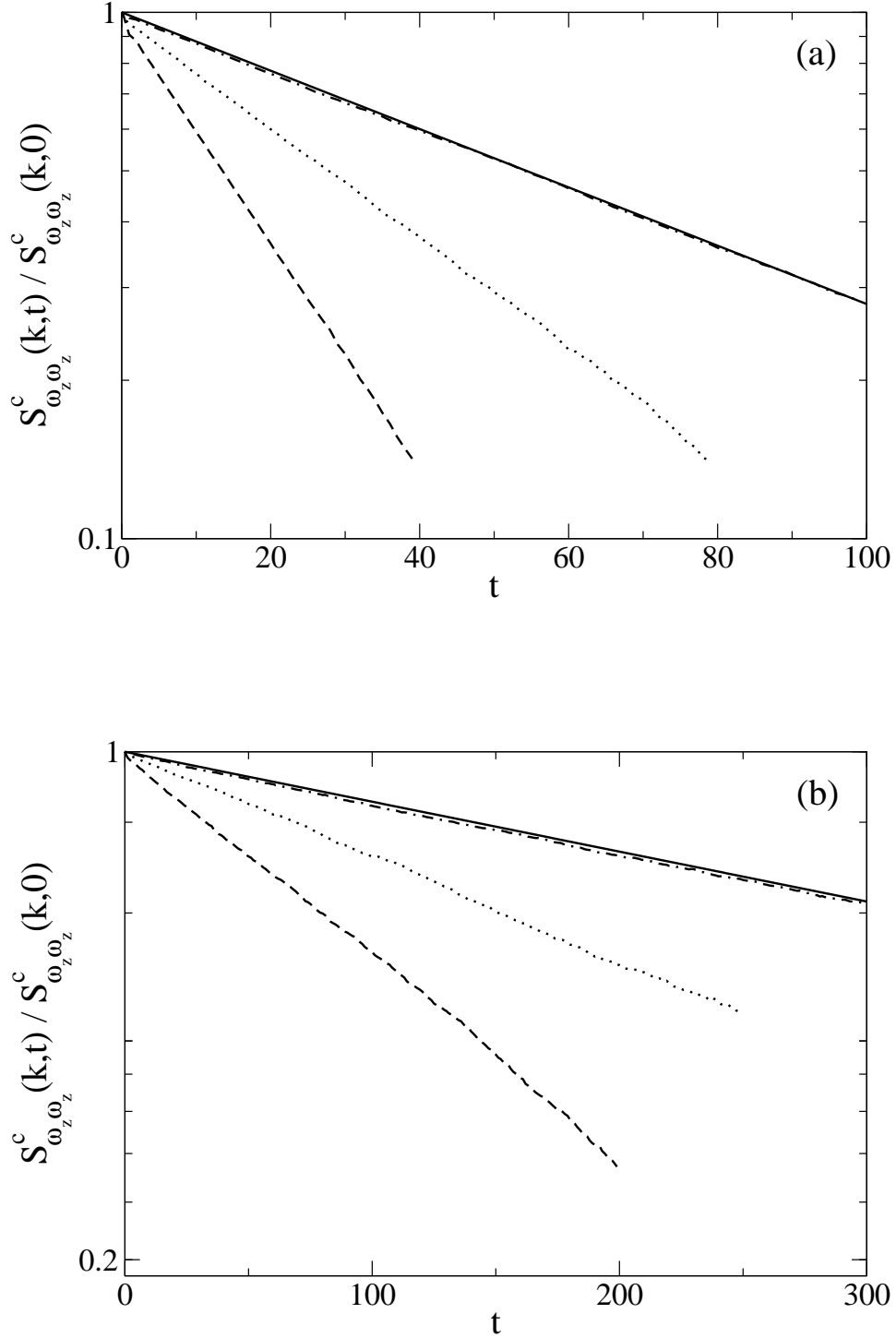

\begin{center}
\vspace{2cm}
\includegraphics[width=5in,angle=0]{vorticity_L32_l1.0_M15_a120_tau1.0_theory.eps}

\vspace{2cm}
\includegraphics[width=5in,angle=0]{vorticity_L32_l0.1_M15_a60_tau1.0_theory.eps}

\caption{Normalized vorticity correlations as a function of time for 
$\bfk=\frac{2\pi}{L}(1,0)$ (dotted-dashed lines), $\bfk=\frac{2\pi}{L}(1,1)$ 
(dotted lines) and $\bfk=\frac{2\pi}{L}(2,0)$ (dashed lines). The solid line 
shows Eq. (\ref{vorticity}) using the theoretical expressions given in Table
\ref{tab_transport} for shear viscosity. The decay profiles were fitted to 
Eq. (\ref{vorticity}) to obtain values for the shear viscosity $\nu$. 
a) $\lambda/a=1.0$ , $\alpha=120^{\circ}$, 
b) $\lambda/a=0.1$, $\alpha=60^{\circ}$. 
Parameters: $L/a=32$, $M=15$ and $\tau=1.0$.
}
\label{fig_vorticity}
\end{center}
\end{figure}

\begin{figure}
\begin{center}
\vspace{2cm}
\includegraphics[width=5in,angle=0]{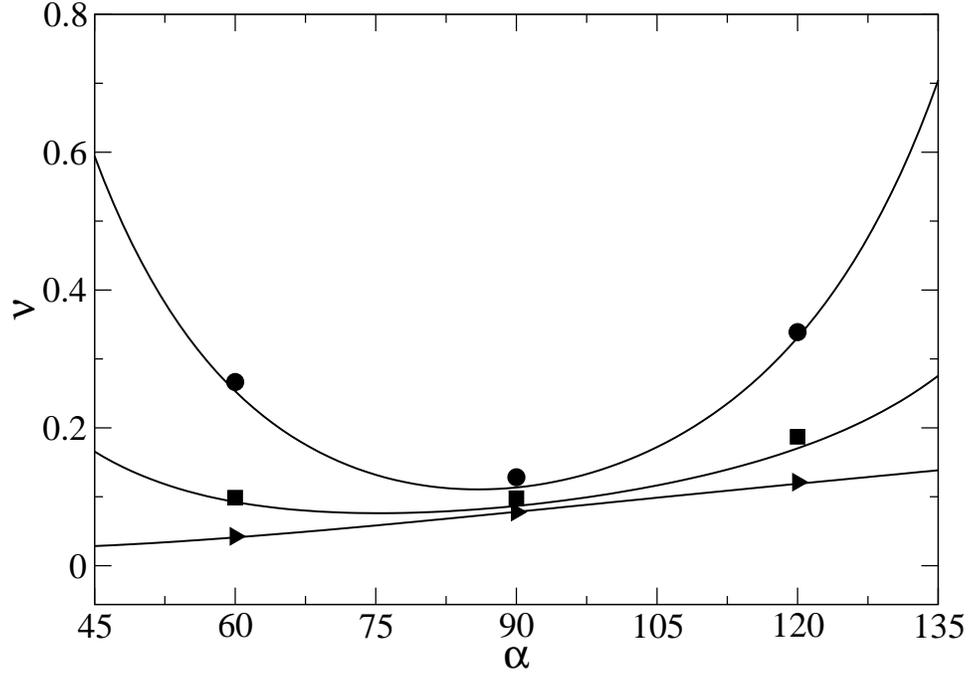}
\caption{Shear viscosity $\nu$ as a function of collision angle $\alpha$ 
for mean free paths $\lambda/a=0.1 ~(\smallblacktriangleright), 
0.5 ~(\smallblacksquare)$, and $1.0 ~(\smallblackcircle)$ measured using 
the decay of the vorticity correlations. The solid lines are the  
theoretical prediction given in Table \ref{tab_transport}, i.e. the sum of 
the kinetic and collisional contributions. 
Parameters: $L/a=32$, $M=15$ and $\tau=1.0$.}
\label{fig_viscosity}
\end{center}
\end{figure}

\noindent{\em Transverse fluctuations}: Instead of the evaluating the 
correlation 
function of the transverse component of the momentum density, it is 
more convenient in simulations to measure the vorticity,  
${\bf w}(\rt)= \nabla \times \bfg_\perp (\rt ) $. 
In two dimensions the vorticity is a scalar, $w_z(\rt)$, and  
the dynamic correlation function decays as
\be
S^c_{w_zw_z}(k,t)=k^2S^c_{g_\perp g_\perp}(k,t) = 
\rho k_BTk^2e^{-\nu k^2 t}.  \label{vorticity}
\ee
Simulation results for the normalized vorticity correlation function for 
$\lambda/a\equiv \tau \sqrt{k_BT/m}=1.0$ with collision angle 
$\alpha=120^\circ$ and $\lambda/a=0.1$ 
with $\alpha=60^\circ$ are shown in Figure \ref{fig_vorticity}.
Here, as with all results presented in this paper, averages are taken over 
$400,000$ iterations and five different random number seeds. The solid 
lines in Figure \ref{fig_vorticity} are a plot of Eq. (\ref{vorticity}) 
using the theoretical prediction for the shear viscosity (sum of kinetic and 
collisional contributions) given in Table \ref{tab_transport}, for the 
smallest wave vector $\bfk=(2\pi/L)(1,0)$. The agreement is excellent. 
We have also fitted the decay profiles for the lowest two wave vectors, 
namely to $\bfk=(2\pi/L)(1,0)$ and $\bfk=(2\pi/L)(0,1)$ with 
Eq. (\ref{vorticity}), and averaged the result to obtain estimates of the 
shear viscosity as a function of collision angle $\alpha$ for different 
mean free paths. The results are presented in Figure \ref{fig_viscosity}, and, 
as expected, the agreement between measured viscosities and the expressions 
given in Table \ref{tab_transport} is very good. These measurements clearly 
show that the theoretical expressions for the shear viscosity are accurate 
even for intermediate mean free paths. 
\bigskip

\begin{figure}
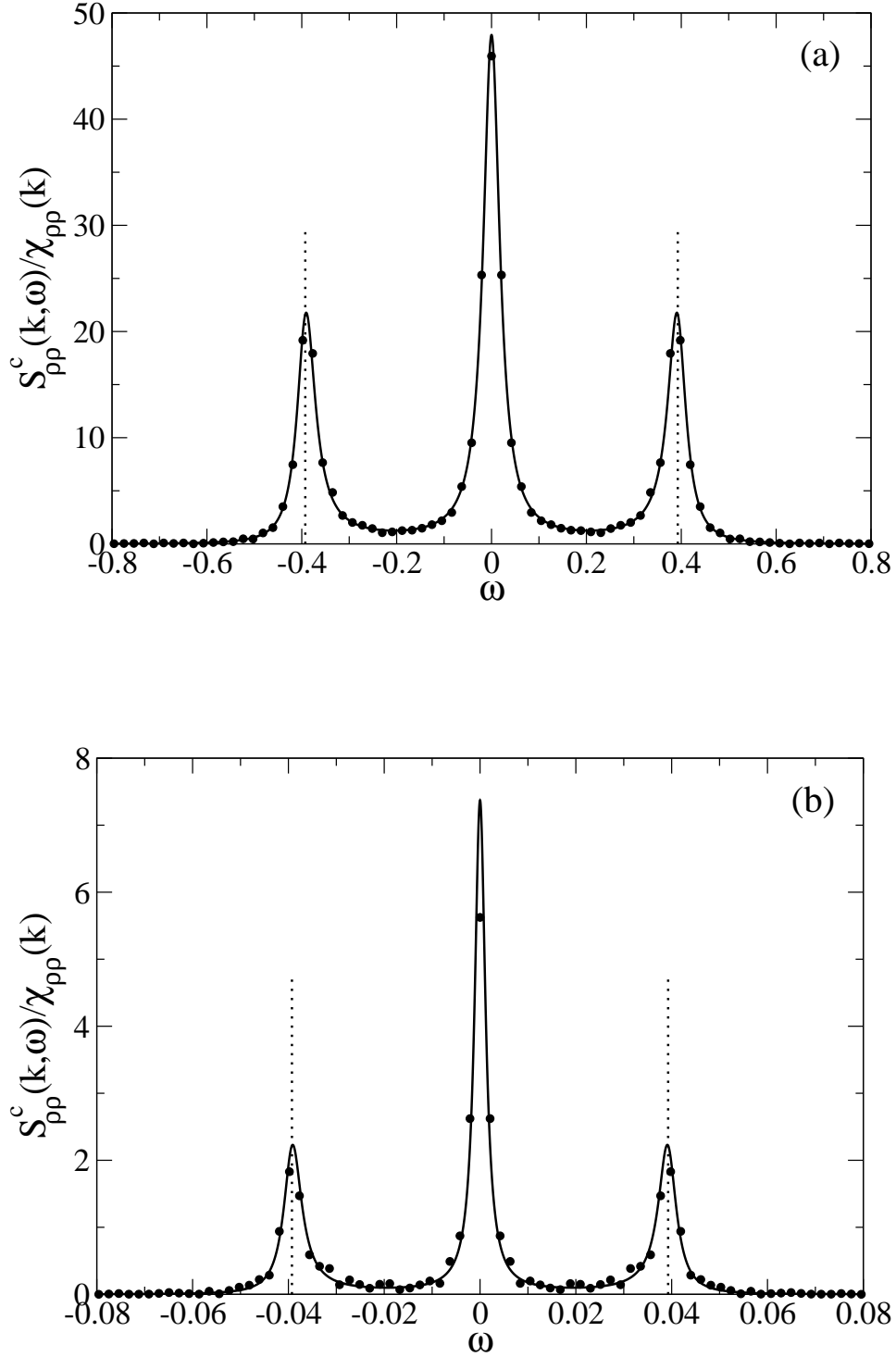

\begin{center}
\vspace{2cm}
\includegraphics[width=5in,angle=0]{structure_L32_l1.0_M15_a120_tau1.0.eps}

\vspace{2cm}
\includegraphics[width=5in,angle=0]{structure_L32_l0.1_M15_a60_tau1.0.eps}

\caption{Normalized dynamic structure factor, 
$S^c_{\rho\rho}(k, \omega)/\chi_{\rho\rho}(k)$, for 
$\bfk=\frac{2\pi}{L}(1,1)$ and a) $\lambda/a=1.0$ with $\alpha=120^{\circ}$, 
and b) $\lambda/a=0.1$ with $\alpha=60^{\circ}$.  The solid line is the 
theoretical prediction obtained using Eq. (\ref{skw}) and the expressions 
for the transport coefficients given in Table \ref{tab_transport}. The dotted 
lines show the predicted positions of the Brillouin peaks using the 
dispersion relation $\omega=\pm ck$. 
Parameters: $L/a=32$, $M=15$ and $\tau=1.0$.}
\label{fig_structure}
\end{center}
\end{figure}

\noindent{\em Longitudinal fluctuations}: Density fluctuations were 
measured at the cell level, and their Fourier-Laplace transform is taken to 
determine the structure factor.  
A naive implementation of this procedure gives wrong results in the 
large frequency region of the spectrum~\cite{inoue_03_dcs}, resulting in 
finite contributions at all frequencies. This problem is well 
known to experimentalists~\cite{tekmen_private}, the solution is to 
first do a Fourier transform to obtain density-density correlations 
as a function of time, symmetrize this result around $t=0$, and then 
perform the Laplace transform from $-\infty$ to $\infty$. Results for the 
structure factor for $\lambda/a=1.0$ with collision angle $\alpha=120^\circ$ 
and $\lambda/a=0.1$ with $\alpha=60^\circ$ are shown in Figures 
\ref{fig_structure}(a) and \ref{fig_structure}(b), respectively. 
The solid lines are the theoretical expression given by Eq. (\ref{skw}) using 
$c=\sqrt{d k_BT/m}$ and values for the transport coefficients obtained using 
the expressions in Table \ref{tab_transport}, assuming that the bulk 
viscosity $\gamma=0$. The agreement is excellent. There are three 
Lorentzian peaks, a central ``Rayleigh peak'' caused by the heat diffusion 
and two symmetrically displaced ``Brillouin peaks'' caused by the sound waves. 
The dotted vertical lines 
in the figures show the theoretically predicted frequencies of the adiabatic 
sound waves in a fluid with an ideal gas equation of state. 

\begin{figure}
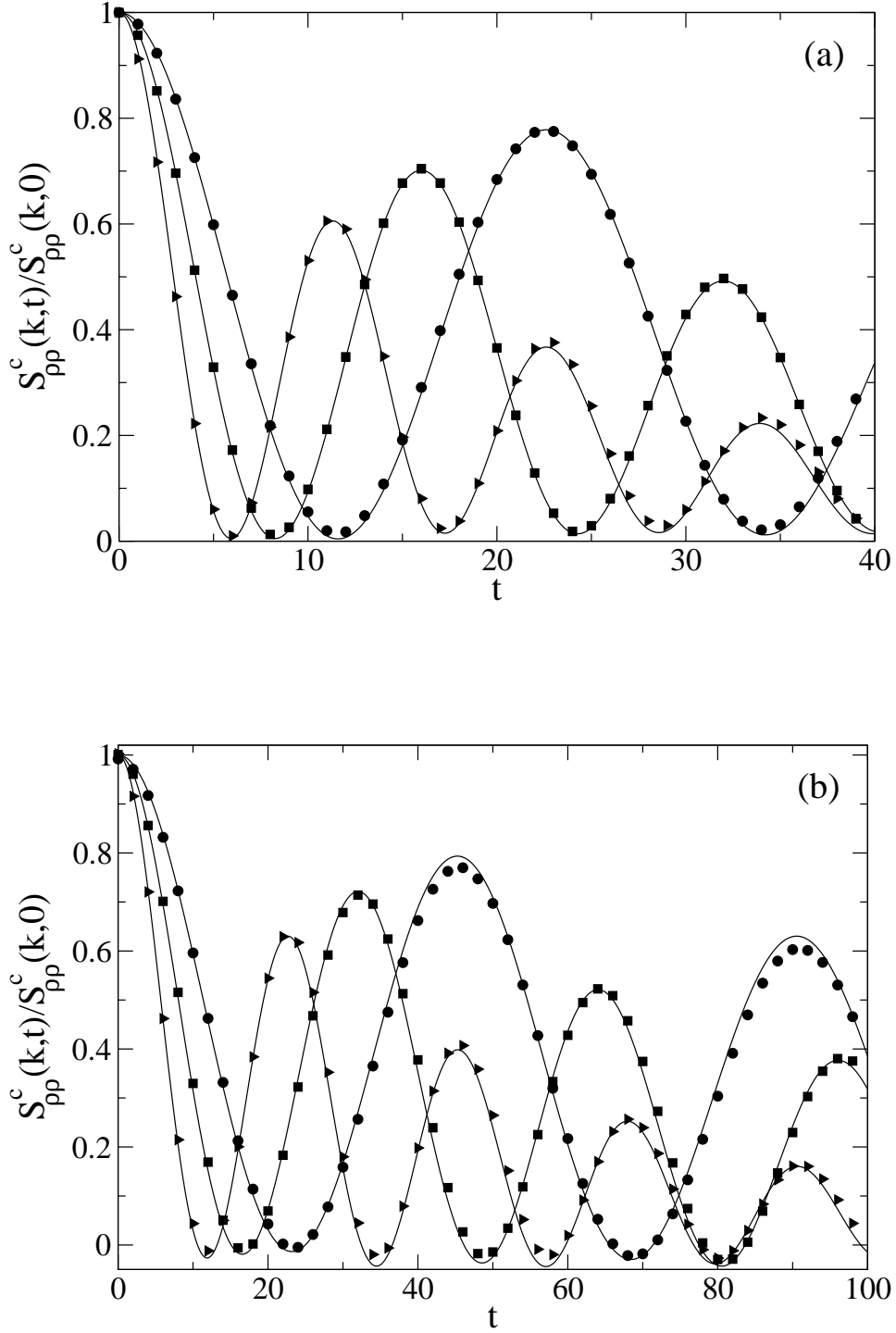

\begin{center}
\vspace{2cm}
\includegraphics[width=5in,angle=0]{density_L32_l1.0_M15_a120_tau1.0.eps}

\vspace{2cm}
\includegraphics[width=5in,angle=0]{density_L32_l0.5_M15_a90_tau1.0.eps}

\caption{Normalized density correlations, 
$S^c_{\rho\rho} (k, t)/S^c_{\rho\rho} (k, 0)$, as a function of time, 
for $\bfk=\frac{2\pi}{L}(1,0)$ ($\smallblackcircle$), 
$\bfk=\frac{2\pi}{L}(1,1)$ ($\smallblacksquare$), and 
$\bfk=\frac{2\pi}{L}(2,0)$ ($\smallblacktriangleright$). 
The solid lines are the theoretical predictions obtained using  
Eq. (\ref{skt}) and the expressions of transport coefficients given in 
Table \ref{tab_transport}. 
a) $\lambda/a=1.0$, $\alpha=120^{\circ}$, and 
b) $\lambda/a=0.5$, $\alpha=90^{\circ}$. 
Parameters: $L/a=32$, $M=15$ and $\tau=1.0$.}
\label{fig_density}
\end{center}
\end{figure}

\begin{figure}
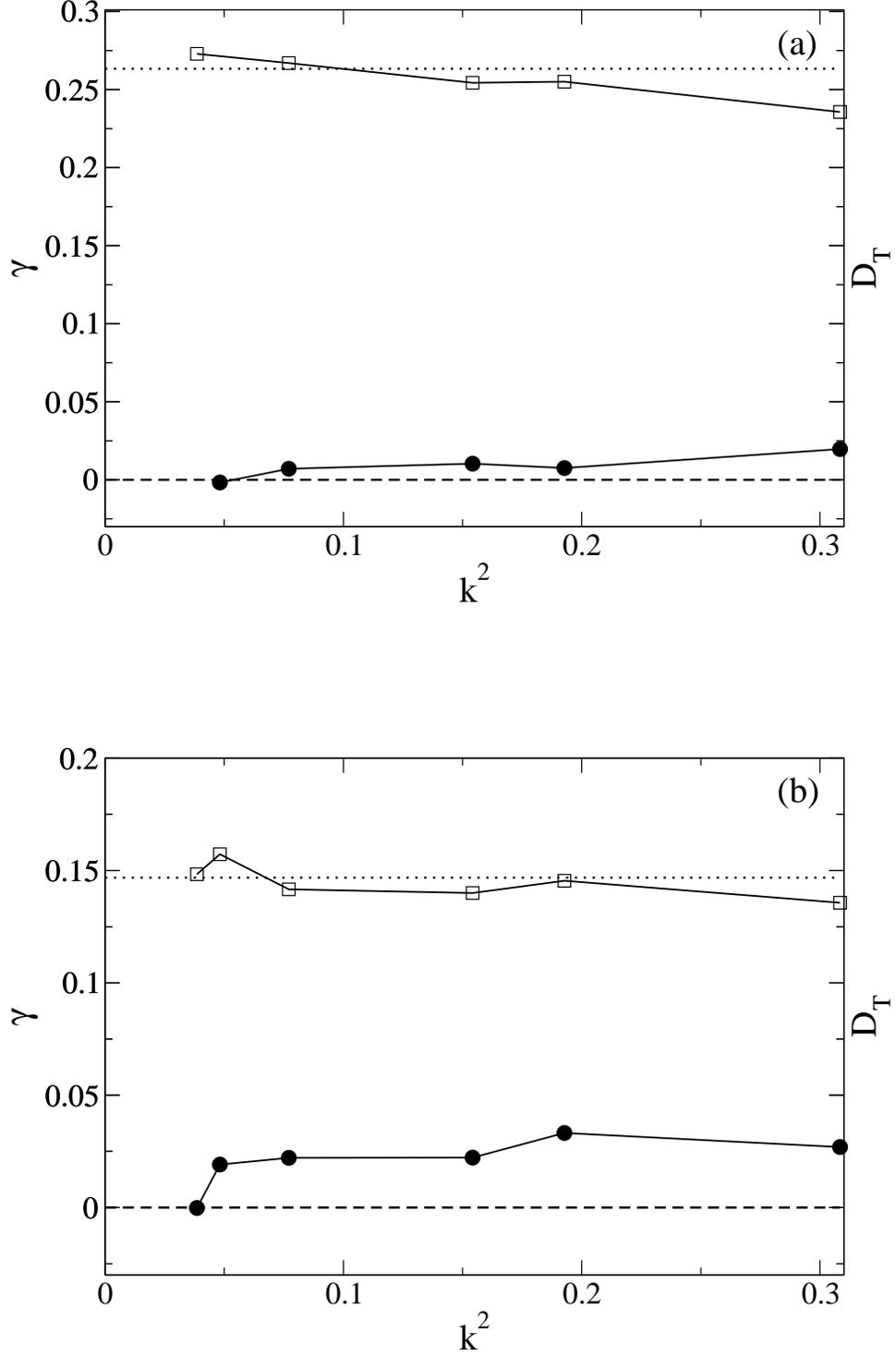

\begin{center}
\vspace{2cm}
\includegraphics[width=5in,angle=0]{bulk_L32_64_l1.0_M15_a120_tau1.0.eps}

\vspace{2cm}
\includegraphics[width=5in,angle=0]{bulk_L32_64_l0.5_M15_a90_tau1.0.eps}

\caption{Fitted values for the bulk viscosity $\gamma$ (\smallblackcircle) 
and thermal diffusivity $D_T$ (\smallwhitesquare) as a function of the wave 
vector squared for a) $\lambda/a=1.0$ and $\alpha=120^{\circ}$, and 
b) $\lambda/a=0.5$ and $\alpha=90^{\circ}$. 
Data was obtained by fitting the time dependent density correlations 
(see Figure \ref{fig_density}) using Eq. (\ref{skt}) while keeping 
$D_T$ and $\gamma$ as free parameters. Dashed and dotted lines represent 
the theoretically predicted values for $\gamma$ and $D_T$, respectively. 
System size $L/a$ ranges from $32$ to $128$. Parameters: $M=15$ and $\tau=1.0$.}
\label{fig_bulk}
\end{center}
\end{figure}

We have also measured time-dependent density correlations for various 
wave vectors.
Figures \ref{fig_density}(a) and \ref{fig_density}(b) contain a comparison 
of the measured time-dependent density correlation functions with the 
predictions of Eq. (\ref{skt}), for $\lambda/a=1.0$ with collision angle 
$\alpha=120^\circ$ and $\lambda/a=0.5$ with $\alpha=90^\circ$. The agreement 
is excellent for all wave vectors considered. The bulk viscosity and 
thermal diffusivity were also independently measured by fitting these time 
dependent density correlations to the form given by Eq. (\ref{skt}) while 
using the theoretically predicted shear viscosity in the sound attenuation 
coefficient, keeping $D_T$ and $\gamma$ as free parameters. The results are 
shown in Figure \ref{fig_bulk} as a function of the wave vector squared, for 
the same set of parameters as in Figure \ref{fig_density}. Once again, the 
theoretical expression for $D_T$ is confirmed, and the bulk viscosity is 
indeed zero.

\begin{figure}
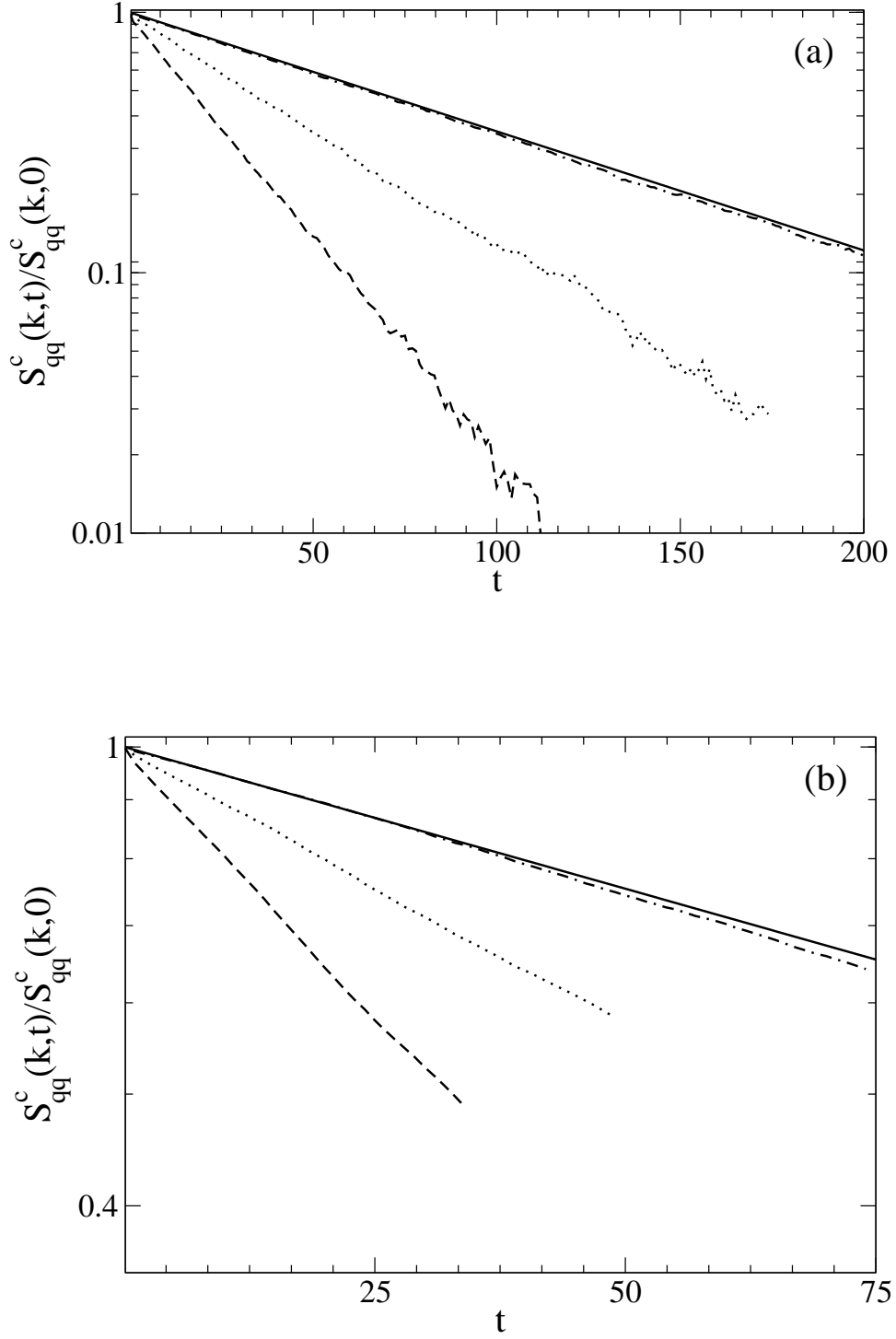

\begin{center}
\vspace{2cm}
\includegraphics[width=5in,angle=0]{cqq_L32_l1.0_M15_a120_tau1.0_theory.eps}
\vspace{2cm}

\includegraphics[width=5in,angle=0]{cqq_L32_l0.5_M15_a90_tau1.0_theory.eps}
\caption{Normalized entropy correlations, $S^c_{qq}(k,t)/S^c_{qq}(k,0)$, 
as a function of time for $\bfk=\frac{2\pi}{L}(1,0)$ (dotted-dashed lines), 
$\bfk=\frac{2\pi}{L}(1,1)$ (dotted lines),  
and $\bfk=\frac{2\pi}{L}(2,0)$ (dashed lines). 
The solid line shows Eq. (\ref{sqq}) using the theoretical expressions 
for the thermal diffusivity, $D_T$,  given in Table \ref{tab_transport}. 
The decay profiles are also fitted to 
Eq. (\ref{sqq}) to obtain unbiased estimates for the thermal diffusivity. 
a) $\lambda/a=1.0$ , $\alpha=120^{\circ}$, 
b) $\lambda/a=0.5$, $\alpha=90^{\circ}$. 
Parameters: $L/a=32$, $M=15$ and $\tau=1.0$.}
\label{fig_cqq}
\end{center}
\end{figure}

\begin{figure}
\begin{center}
\vspace{2cm}
\includegraphics[width=5in,angle=0]{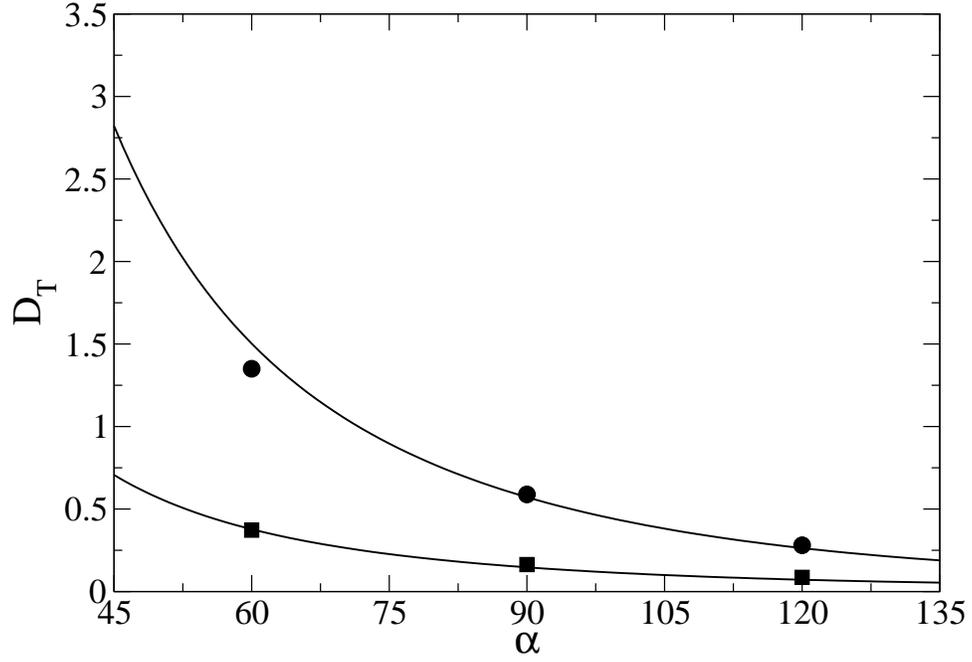}
\caption{Estimates for the thermal diffusivity $D_T$  
as a function of collision angle $\alpha$ for mean free paths 
$\lambda/a=0.5 ~(\smallblacksquare)$ and 
$1.0 ~(\smallblackcircle)$ obtained by fitting the decay of entropy 
correlations, $S^c_{qq}(\bfk, t)$. The solid lines are the theoretical 
prediction given in Table \ref{tab_transport}, i.e. the sum of kinetic 
and collisional contributions. Parameters: $L/a=32$, $M=15$ and $\tau=1.0$.}
\label{fig_DT}
\end{center}
\end{figure}

In order to obtain an independent measure of the thermal diffusivity, we have 
measured the temporal behavior of the entropy correlations, $S_{qq}^c(k,t)$. 
The results are shown in Figure \ref{fig_cqq} for $\lambda/a=1.0$ with 
collision angle $\alpha=120^\circ$ and $\lambda/a=0.5$ with $\alpha=90^\circ$. 
As expected, these corrrelations decay exponentially for all wave vectors 
considered. The solid lines in Figure \ref{fig_cqq}
are a plot of Eq. (\ref{sqq}) for the smallest wave vector $\bfk=(2\pi/L)(1,0)$, 
using the theoretical prediction for the total thermal diffusivity, 
$D_T$ (see Table \ref{tab_transport}), and the agreement is again very good. 
As was done for the vorticity measurements, we have also fitted the decay 
profiles for the lowest two wave vectors with Eq. (\ref{sqq}), and averaged
the results to obtain independent measurements of the thermal diffusivity.
The results for these measurements are shown in Figure \ref{fig_DT} as a 
function of the collision angle $\alpha$ for $\lambda/a=0.5$ and 
$\lambda/a=1.0$. The theoretical values obtained using the formulae for 
$D_T$ in Table \ref{tab_transport} (sum of kinetic and collisional 
contributions) are shown in solid lines. These results are the first direct 
equilibrium measurements of the thermal diffusivity. 

Finally, it is important to emphasize that just as for the shear viscosity, 
collisional contributions provide the dominant contribution to the thermal 
diffusivity at small mean free path. Figure \ref{fig_nuDT_compare} shows 
the theoretically predictions for both the collisional and kinetic 
contributions to the shear viscosity and thermal diffusitivity (inset) as 
a function of the mean free path $\lambda/a$. Collisional contributions to 
both transport coefficients are particularly important for small mean free 
paths and small $M$.

\section{Correlation effects}

Green-Kubo relations for the SRD transport coefficients have been derived 
in Ref. \cite{ihle_03_srd_a} and analyzed in Refs. \cite{ihle_03_srd_b,
tuzel_03_tcs,ihle_04_rgr,ihle_05_ect}, where it was shown that there are 
both kinetic and collisional contributions to the shear viscosity and the 
thermal diffusivity. The collisional contributions to these transport 
coefficients have been discussed in detail in Ref. \cite{ihle_05_ect}. 

The kinetic contribution to the transport coefficients have been derived 
by several groups  
\cite{malevanets_99_mms,ihle_03_srd_b,tuzel_03_tcs,ihle_04_rgr,pooley_05_ktd} 
assuming molecular chaos. The results of these calculations are summarized 
in Table \ref{tab_transport}. Simulation results for the shear viscosity and 
thermal diffusivity have generally been found to be in good agreement with 
these predictions. However, it is known that there are 
correlation effects for $\lambda/a$ smaller than one \cite{ihle_04_rgr}. 
They arise from correlated collisions between particles that are in the 
same collision cell for more than one time step. In the following, we expand 
on the discussion of Ref. \cite{ihle_04_rgr} and calculate the first 
correlation corrections to both the viscosity and the self-diffusion 
coefficient explicitly. Similar calculations can in principle be done for the correlation contributions 
to thermal diffusivity. The reason that these corrections to $\nu$ and 
$D_T$ are generally negligible is that they are only significant in the 
small $\lambda/a$ regime, where the collisional contribution to the 
transport coefficients dominates. 

\begin{figure}
\begin{center}
\vspace{2cm}
\includegraphics[width=5in,angle=0]{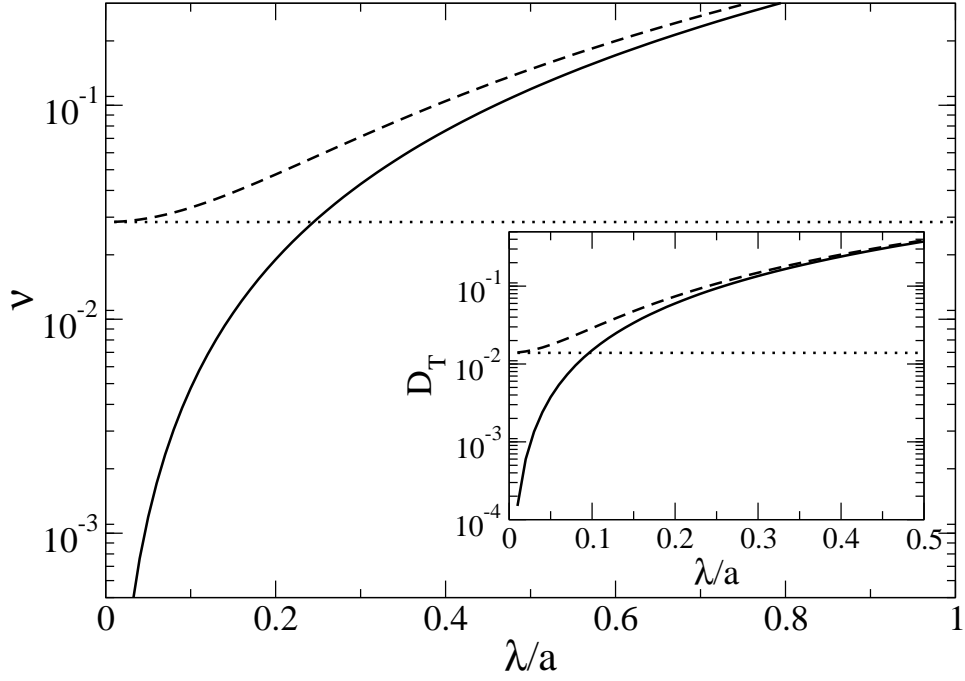}
\caption{Shear viscosity $\nu$ and thermal diffusivity $D_T$ (inset) as a 
function of $\lambda/a$. Both plots are obtained using the theoretical 
expressions given in Table \ref{tab_transport}. The solid and dotted lines 
are the kinetic and collisional contributions, respectively. The dashed 
lines are the total contributions to these transport coefficients. For 
consistency, in the calculation of thermal diffusivity, both the kinetic 
and collisional contributions are taken only up to and including $O(1/M)$.
In the plots, $\lambda/a$ was varied by changing $k_BT$ for 
fixed $\tau=1.0$. Parameters: $M=3$ and $\alpha=60^\circ$.} 
\label{fig_nuDT_compare}
\end{center}
\end{figure}

Figure \ref{fig_nuDT_compare} shows a comparison of 
both kinetic and collisional contributions to shear viscosity and thermal diffusivity (inset) for $M=3$ and
$\alpha=60^\circ$.
Correlation effects would be 
most pronounced when both contributions are comparable, i.e. at a mean free 
path of $\lambda/a\simeq 0.25$ for $\nu$ and 
$\lambda/a\simeq 0.1$ for $D_T$ (see Figure \ref{fig_nuDT_compare}).
On the other hand, because there are 
no collisional contributions to the self-diffusion coefficient, correlation 
corrections dramatically increase the value of this transport coefficient 
in the small $\lambda/a$ regime. It is important to note that there are 
no correlation corrections to the collisional contributions so that the 
expressions for the collisional shear viscosity 
\cite{ihle_04_rgr,ihle_05_ect,pooley_05_ktd} and collisional thermal diffusivity 
\cite{ihle_04_rgr,ihle_05_ect} are exact. 

In the following, we restrict ourselves to two dimensions; the same analysis, 
however, can also be used in three dimensions. Expressions for the 
shear viscosity and the self-diffusion coefficient in this section obtained in 
the molecular chaos approximation will 
include contributions from fluctuations in the number of particles per cell. 
However, when calculating correlation corrections, we will assume that 
the number of particles per cell, $M$, is fixed. Including these fluctuations 
is straightforward but tedious, and since it would not provide 
any additional insight into the underlying phenomena, we have decided to ignore 
this effect in the following. 

\subsection{Shear viscosity}

The Green-Kubo relation for the kinetic contribution to the shear viscosity 
\cite{ihle_04_rgr} is
\begin{equation}\label{nu_kin}  
\nu^{kin}=  
k_{B}T\tau\sum_{n=0}^{\infty}{^\prime} G (n\tau) 
\end{equation}
where the prime indicates that the $t=0$ contribution to the sum occurs with 
the weight $1/2$, and  
\begin{equation}\label{G_ntau}
G(n\tau) = \sum_{i,j}\langle v_{ix}(0)v_{iy}(0) v_{jx}(n\tau)v_{jy}(n\tau) 
\rangle  /N(k_{B}T)^{2}\;.
\end{equation}

In the molecular chaos approximation~\cite{ihle_04_rgr}, 
\begin{equation}
G(n\tau)\equiv G_{c}(n\tau)=\left[1-2\sin^{2}\left(\alpha\right)
\left(M-1+e^{-M}\right)/M\right]^{n}\;. \label{Gc}
\end{equation}
Inserting this expression into Eq. (\ref{nu_kin}) and summing, one obtains 
the kinetic contribution to viscosity given in Table \ref{tab_transport}. 
Figure \ref{fig_sigma_comp} contains a comparison of simulation results 
for $G(n\tau)$ with the molecular chaos approximation Eq. (\ref{Gc}). 
As can be seen, the first and the most important correlation contribution 
to $\nu^{kin}$ occurs for $n=2$. The functional form of this leading 
correlation correction, 
$\delta G(2\tau) = G(2\tau) - G_{c}(2\tau)$, can be calculated analytically. 

\begin{figure}
\begin{center}
\vspace{2cm}
\includegraphics[width=2.5in,angle=0]{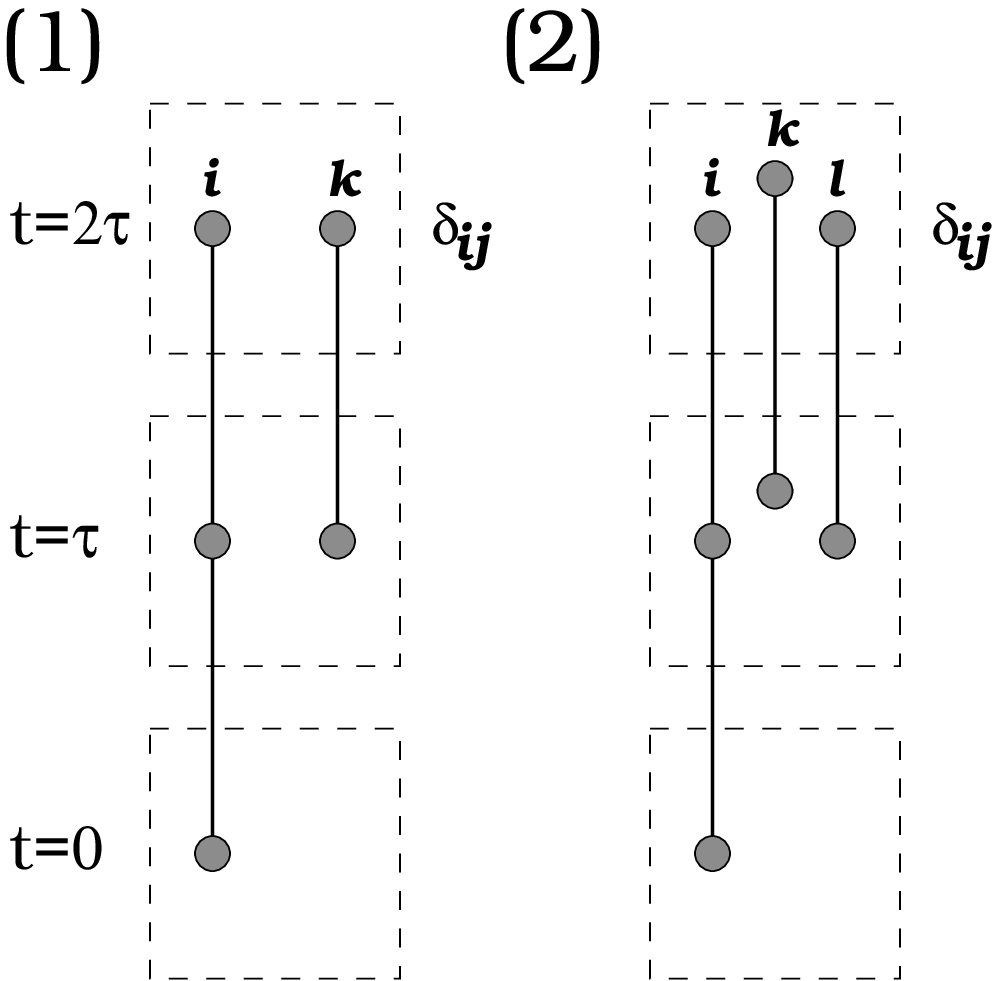}

\vspace{0.8cm}
\includegraphics[width=2.5in,angle=0]{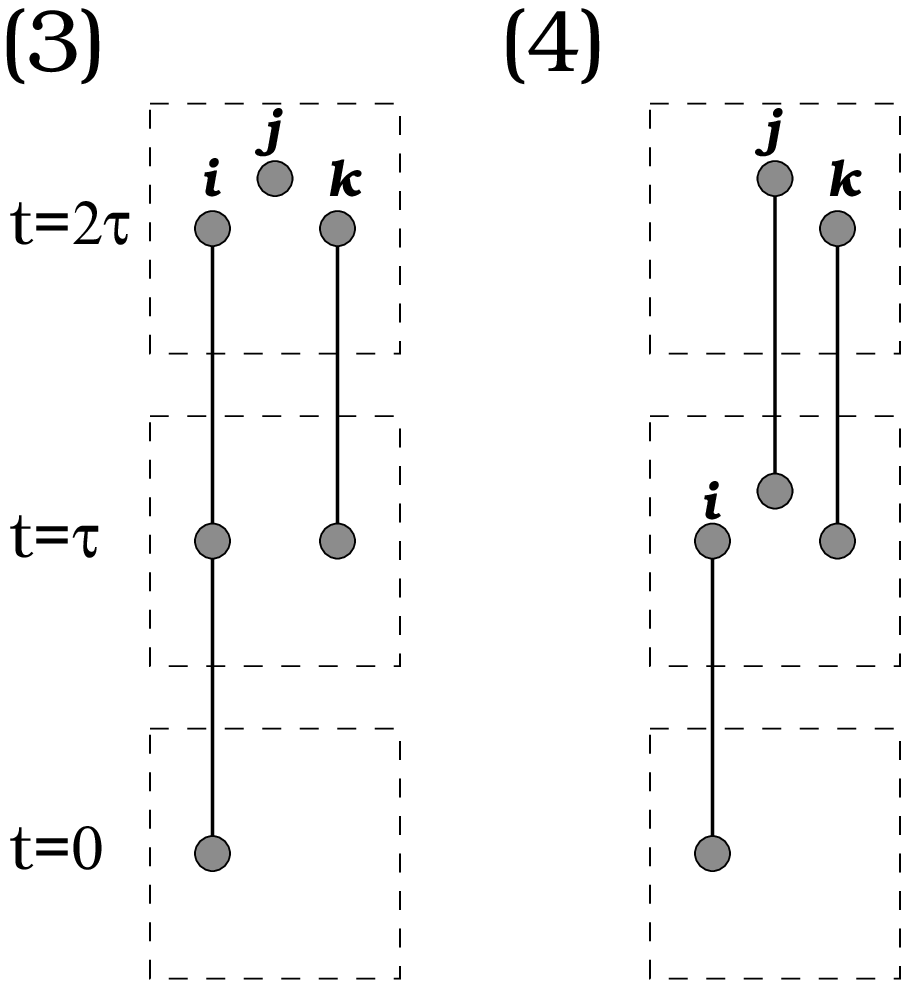}
\vspace{0.8cm}

\includegraphics[width=2.5in,angle=0]{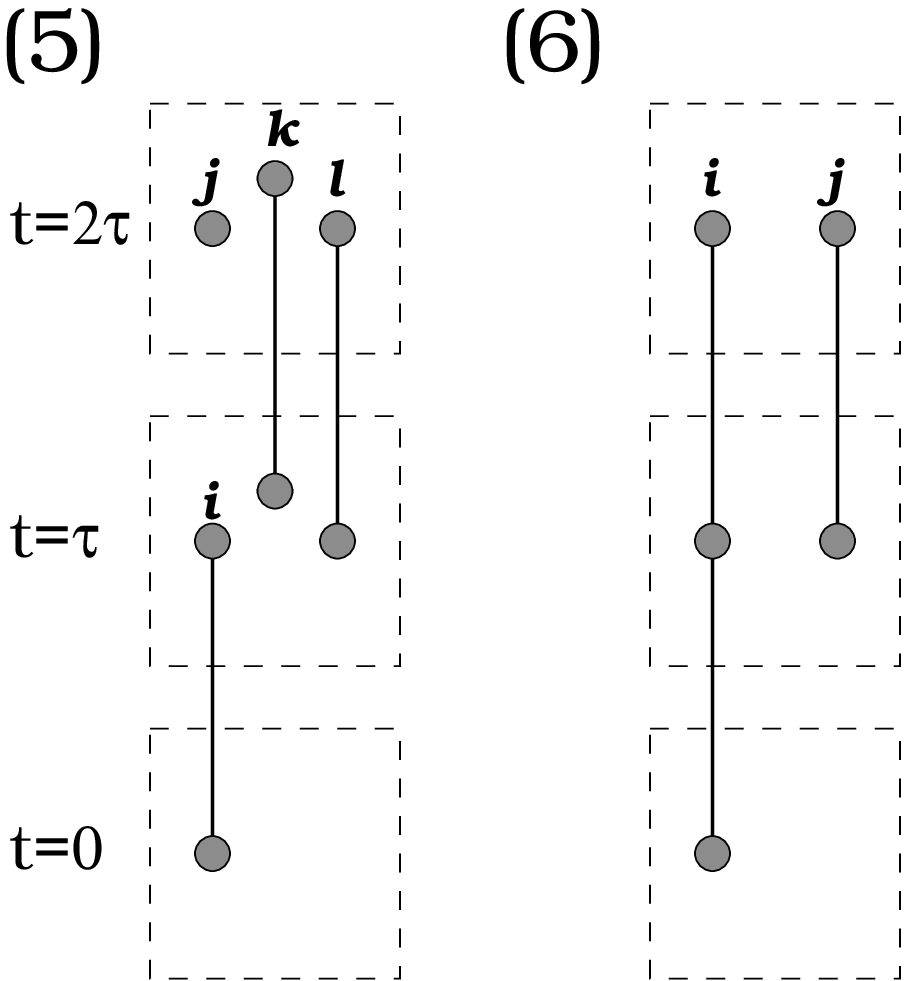}
\caption{Diagrams contributing to correlations at $t=2\tau$ in the 
calculation of kinetic contributions to shear viscosity. The first two 
diagrams show the diagonal and the others show the off-diagonal contributions.}
\label{fig_correlation}
\end{center}
\end{figure}

As illustrated schematically in Figure \ref{fig_correlation}, there are 
six distinct particle configurations which contribute. The first two, shown 
in Figures \ref{fig_correlation}(1) and \ref{fig_correlation}(2)  
occur when $i=j$ in the sum in Eq. (\ref{G_ntau}); we will call these the 
diagonal contribution. In Figure \ref{fig_correlation}(1), particle $k$ is 
in the same collision cell as $i$ for both $t=2\tau$ and $t=\tau$. 
In Figure \ref{fig_correlation}(2), two distinct particles, labeled $k$ and 
$l$, are in the same collision cell as $i$ at both $t=2\tau$ and $t=\tau$.  
Other (off-diagonal) contributions, which occur for $i\ne j$, 
are given in Figure \ref{fig_correlation}(3)-(6). 
These contributions are significant only at small mean free paths, since 
their amplitudes are proportional to the probability 
that two or more particles are in the same collision cell 
for multiple times. 


\subsubsection{Diagonal contributions}

The first diagonal contribution, which we denote by  
$\delta G_{1}(2\tau)$, occurs when two particles, with indices $i$ and $k$, 
are in the same collision cell at both $t=\tau$ and $t=2\tau$. The probability 
for this to occur is $p_{2}$; $p_2$ is calculated in the $\lambda/a\to0$ 
limit in Appendix A. 
In two dimensions the velocity of particle $i$ at time $t=n\tau$ is related 
to the velocities of its collision partners at $t=(n-1)\tau$ by
\begin{eqnarray}\label{collision}
v_{ix}(n\tau)&=&c v_{ix}([n-1]\tau)+\frac{1-c}{M}\sum_{k} v_{kx}([n-1]\tau)+
s\left[v_{iy}([n-1]\tau)-\frac{1}{M}\sum_{k}v_{ky}([n-1]\tau)\right] \\
v_{iy}(n\tau)&=&c v_{iy}([n-1]\tau)+\frac{1-c}{M}\sum_{k} v_{ky}([n-1]\tau)-
s\left[v_{ix}([n-1]\tau)-\frac{1}{M}\sum_{k}v_{kx}([n-1]\tau)\right] \;,
\nonumber
\end{eqnarray}
where $c=\cos(\alpha)$ and $s=\sin(\alpha)$. Using Eq. (\ref{collision}) to 
relate the velocities at $t=2\tau$ to those at $t=\tau$, we have  
\begin{equation}\label{g1} 
\delta G_{1}(2\tau)=2 p_2 (M-1)\sum_{i}\left\langle v_{ix}(0)v_{iy}(0)
\left[\z_{1}v_{ix}(\tau)+\z_{2}v_{iy}(\tau)\right]
\left(\frac{1-c}{M}v_{ky}(\tau)+\frac{s}{M}v_{kx}(\tau)\right) \right\rangle 
/N(k_{B}T)^{2}, 
\end{equation}
where $\zeta_1=1/M+c(1-1/M)$ and $\zeta_2=s(1-1/M)$, and 
the factor $(M-1)$ accounts for the sum over $k\ne i$ and the factor 
2 comes from the fact that $i$ and $k$ can interchange roles. 
The equilibrium average in Eq. (\ref{g1}) entails the average over all initial 
coordinates and velocities (at $t=0$) as well as averages over the stochastic 
rotations ($\pm\alpha$) at $t=\tau$ and $t=2\tau$. 
Performing the average over the collision angle at $t=2\tau$ in Eq. (\ref{g1}) 
removes all terms linear in $s$, so that 
\begin{equation}
\delta G_1(2\tau)=\frac{2 p_2 (M-1)}{M}\sum_{i}\langle v_{ix}(0)v_{iy}(0) 
[\z_{1}(1-c)v_{ix}(\tau)v_{ky}(\tau)+\z_{2}s v_{iy}(\tau)v_{kx}(\tau)]
\rangle /N(k_{B}T)^2. 
\end{equation}
Using Eq. (\ref{collision}) once again with $n=1$,
\begin{eqnarray}
\delta G_1(2\tau)=\frac{2 p_2  (M-1)}{M}\sum_i\langle 
v_{ix}(0)v_{iy}(0)&& \!\!\!\!\!\!\!\!\!\!\!\left[\z_{1}(1-c)
[\z_{1}v_{ix}(0)+\z_{2}v_{iy}(0)]
\left( \frac{1-c}{M}v_{iy}(0)+\frac{s}{M}v_{ix}(0)\right)\right.
\nonumber \\
&+& \left. \z_{2} s [\z_{1}v_{iy}(0)-\z_{2}v_{ix}(0)]
\left( \frac{1-c}{M}v_{ix}(0)-\frac{s}{M}v_{iy}(0)\right)\right]\rangle 
/N(k_{B}T)^2 .  
\end{eqnarray}
Averaging now over the collision angle at $t=\tau$ and the particle velocities 
and coordinates at $t=0$ yields 
\begin{equation}
\delta G_{1}(2\tau)=\frac{2p_2  (M-1)}{M^{2}}\left[ \z_{1}(1-c)+\z_{2} s 
\right]^{2} \;.
\end{equation}

The other diagonal contribution, which we will denote as $\delta G_{2}(2\tau)$, 
arises when three particles, with indices $i$, $k$ and $l$, are in the same 
collision cell at $t=\tau$ and $t=2\tau$. The probability that three 
particles are in the same collision cell in consecutive time steps will 
be denoted by $p_3$; $p_3$ is calculated in the $\lambda/a\to0$ limit in 
Appendix B. Using Eq. (\ref{collision}),  
\begin{equation}
\delta G_{2}(2\tau)=p_3 (M-1)(M-2)\sum_{i}\left\langle v_{ix}(0)v_{iy}(0)
\left(\frac{1-c}{M}v_{kx}(\tau)-\frac{s}{M}v_{ky}(\tau)\right)
\left(\frac{1-c}{M}v_{ly}(\tau)+\frac{s}{M}v_{lx}(\tau)\right) 
\right\rangle /N(k_{B}T)^2,
\end{equation}
where the prefactor $(M-1)(M-2)$ comes from the double sum over $k,l$. 
Averaging over the collision angle at $t=2\tau$ yields
\begin{equation}
\delta G_{2}(2\tau)=\frac{p_3(M-1)(M-2)}{M^{2}}\sum_{i}\left\langle 
v_{ix}(0)v_{iy}(0)\left([1-c]^2 v_{kx}(\tau)v_{ly}
(\tau)-s^{2}v_{ky}(\tau)v_{lx}(\tau)\right) \right\rangle /N(k_{B}T)^{2}.
\end{equation}
Finally, using Eq. (\ref{collision}) again and averaging over 
the collision angle at $t=\tau$ and velocities and coordinates at $t=0$ yields  
\begin{equation}
\delta G_{2}(2\tau)=\frac{p_3(M-1)(M-2)}{M^4}\left[2c(c-1)\right]^2 \;.
\end{equation}

\subsubsection{Off-diagonal contributions}

The analysis of these contributions is very similar to that which was used 
to evaluate the diagonal contributions, so we provide fewer details 
than in the previous subsection. There are four off-diagonal contributions, 
all of which contribute with probability $p_2$. 
The first, $\delta G_{3}({2\tau})$, shown in Figure 
\ref{fig_correlation}(3), can be written as 
\begin{equation}
\delta G_{3}(2\tau)=2p_2(M-1)(M-2)\sum_{i}
\left\langle v_{ix}(0)v_{iy}(0)\left(\frac{1-c}{M}v_{ix}(\tau)-
\frac{s}{M}v_{iy}(\tau)\right)
\left(\frac{1-c}{M}v_{ky}(\tau)+\frac{s}{M}v_{kx}(\tau)\right) 
\right\rangle /N(k_{B}T)^{2}\;,
\end{equation}
where the factor $2(M-1)(M-2)$ accounts for $i$ and $k$ interchanging roles 
and the double sum over $j$ and $k$. Following the same procedure as was 
used to evaluate $\delta G_1(2\tau)$ and $\delta G_2(2\tau)$, we have 
\begin{equation}
\delta G_{3}(2\tau)=\frac{2p_2(M-1)(M-2)}{M^3}
[2c(c-1)][ \z_{1}(1-c)+\z_{2}s] \;.
\end{equation}
The second contribution, $\delta G_{4}({2\tau})$, depicted Figure 
\ref{fig_correlation}(4) can be written as
\begin{equation}
\delta G_{4}(2\tau)=2p_2(M-1)(M-2)\sum_{i}
\left\langle v_{ix}(0)v_{iy}(0)[\z_{1}v_{jx}(\tau)+\z_{2}v_{jy}(\tau)]
\left(\frac{1-c}{M}v_{ky}(\tau)+\frac{s}{M}v_{kx}(\tau)\right) 
\right\rangle /N(k_{B}T)^2,
\end{equation}
where the prefactor is similar to that of $\delta G_{3}(2\tau)$.
Using this result, it is straightforward to show that  
\begin{equation}
\delta G_{4}(2\tau)=\frac{2p_2(M-1)(M-2)}{M^3}
[2c(c-1)][\z_{1}(1-c)+\z_{2}s] \;.
\end{equation}
The third contribution involves the configuration shown in Figure 
\ref{fig_correlation}(5). It is similar to the diagonal contribution 
$\delta G_{2}(2\tau)$ except for the fact that the particle with index $i$ 
is not  in the same shifted cell as $k$ and $l$ at time $t=2\tau$. Only 
two particles are in the same collision cell for consecutive time steps, 
so that the relevant probability is $p_2$. By analogy with the 
expression for $\delta G_{2}(2\tau)$, 
\begin{equation}
\delta G_{5}(2\tau)=\frac{p_2(M-1)(M-2)(M-3)}{M^4}
[2c(c-1)]^2 \;,
\end{equation}
where the factor of $(M-3)$ comes from the additional sum over $j$ with 
the constraints $i\neq j\neq k\neq l$.

\begin{figure}
\begin{center}
\vspace{2cm}
\includegraphics[width=5in,angle=0]{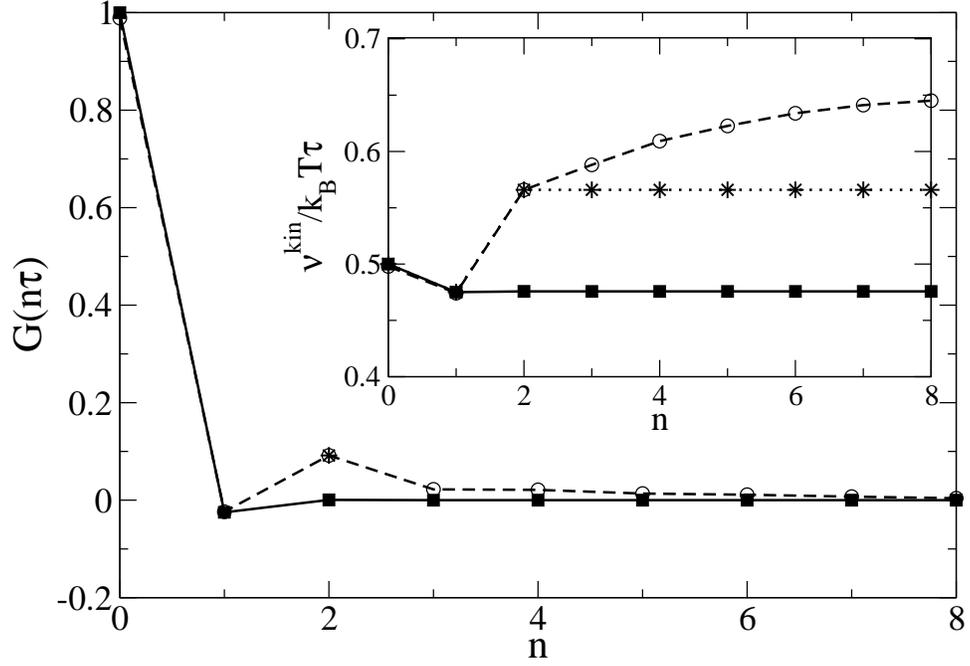}
\caption{Stress correlations, $G(n\tau)$, as a function of time step $n$ 
for $\lambda/a=0.25$ and $M=3$. The inset shows the normalized kinetic 
contribution to shear viscosity, $\nu^{kin}/k_BT\tau$, as a function of 
time step. The measured values are open circles ($\smallwhitecircle$), and 
the results of molecular chaos approximationi, $G_c(n\tau)$ 
(see Eq. (\ref{Gc})), are filled squares ($\smallblacksquare$). The asterisk 
($\smallstar$) in the main figure is the prediction of Eq. (\ref{deltaGsum}) 
using the numerically determined values of the probabilities. The asterisks 
in the inset are a plot of Eq. (\ref{nusum}). 
Parameters: $L/a=64$, $\alpha=60^\circ$, $\tau=1.0$.} 
\label{fig_sigma_comp}
\end{center}
\end{figure}

\begin{figure}
\begin{center}
\vspace{1.7cm}
\includegraphics[width=5in,angle=0]{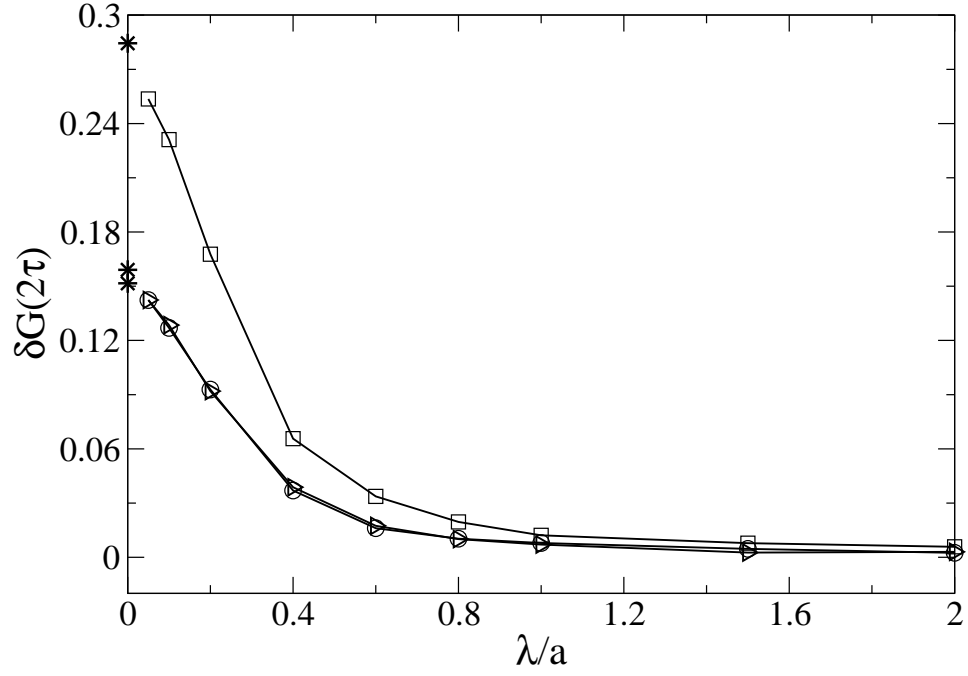}
\caption{Correlation contributions $\delta G(2\tau)$ as a function of 
$\lambda/a$. Results for collision angles $\alpha=60^{\circ}$ 
($\smallwhitecircle$), $90^{\circ}$ ($\smallwhitesquare$),  
and $120^{\circ}$ ($\smallwhitetriangleright$) are presented. 
The asterisks ($\smallstar$) are the 
theoretical predictions in the limit $\lambda/a\rightarrow 0$.  
Parameters: $L/a=64$, $M=5$, $\tau=1.0$.}
\label{fig_G}
\end{center}
\end{figure}

The final contribution, $\delta G_{6}(2\tau)$, occurs when $i$ and $j$ 
are in the same collision cell at both $t=2\tau$ and $t=\tau$ (see 
Figure \ref{fig_correlation}(6)). This contribution is given by 
\begin{equation}
\delta G_{6}(2\tau)=2p_2(M-1)\sum_{i}\left\langle v_{ix}(0)v_{iy}(0)
\left(\frac{1-c}{M}v_{ix}(\tau)-\frac{s}{M}v_{iy}(\tau)\right) 
[\z_{1}v_{jy}(\tau)-\z_{2}v_{jx}(\tau)]\right\rangle /N(k_{B}T)^2,
\end{equation}
where the factor $2(M-1)$ accounts for the sum over $j$ and interchanging 
$i$ and $j$.
Following the same procedure as for the other contributions, we find 
\begin{equation}
\delta G_{6}(2\tau)= \frac{2p_2(M-1)}{M^2}
[\z_{1}(1-c)+\z_{2} s] \;.
\end{equation}
The total correlation enhancement is obtained by summing these six 
contributions, 
\begin{equation}
\delta G(2\tau)= \sum_{n=1}^6  \delta G_n(2\tau) \;.
\label{deltaGsum}
\end{equation}
Note that the only dependence on the temperature and time step $\tau$ 
occur in the probabilities $p_2$ and $p_3$. The measured correlation 
contributions to the shear viscosity are shown in Figure \ref{fig_G} as a 
function of $\lambda/a$ for $\alpha=60^\circ$, $90^\circ$, and $120^\circ$. 
Using the $\lambda/a\rightarrow0$ values for the probabilities $p_2$ and $p_3$ 
calculated in Appendices A and B,
one finds 
\begin{equation}
\delta G(2\tau)= \sum_{n=1}^6  \delta G_n(2\tau) = \frac{M-1}{9M^4}
\left[\{4M[1+\cos (\alpha)]\}^2+7(2-M)\cos^2 (\alpha) \right] \;.
\label{deltaG}
\end{equation}
and 
\begin{equation}
\nu^{kin}\simeq k_BT\tau \left\{ \sum_{n=0}^{\infty}{^\prime} G_c (n\tau) 
+\delta  G(2\tau) \right\} \label{nusum}
\end{equation}
for the shear viscosity. As can be seen from Figure \ref{fig_G}, the 
results obtained using Eq. (\ref{deltaG}) (asterisks) are in excellent 
agreement with simulation data in the limit of zero mean free path. 
More generally, we have determined the probabilities $p_2$ and $p_3$ 
numerically and have found that they depend only on the value of the 
mean free path, $\lambda$, and not on $\tau$ and $T$ individually 
(see inset of Figure \ref{fig_H} for a plot of $p_2$ as a function of 
$\lambda/a$). Finally, these results can be used in Eq. (\ref{deltaGsum}) 
to obtain an estimate of the correlation contribution $\delta G(2\tau)$ 
for arbitrary $\lambda$. The asterisks in Figure \ref{fig_sigma_comp} show 
the results of this procedure, and as might be expected, the agreement is 
excellent. Finally, simulation results for the total kinetic contribution 
to the viscosity as a function of time are shown in the inset to Figure 
\ref{fig_sigma_comp}. The filled squares ($\smallblacksquare$) are the 
predictions of molecular chaos approximation, and asterisks are a plot 
of Eq. (\ref{nusum}). The incipient long-time tail is clearly visible
in the figure. This is one of the reasons that it has been difficult to obtain
good estimates for the ``bare'' kinetic contribution to the transport 
coefficients. In principle, the methods used to determine $\delta G(2\tau)$ 
can also be employed to determine these correlation contributions at greater 
time lags. The corresponding probabilities, however, that particles become 
collision neighbors after a finite time interval are much harder to 
determine, since they depend in detail to relative probabilities of various 
fluctuating flow configurations. 

\subsection{Self-diffusion coefficient}

The Green-Kubo relation for the self-diffusion coefficient is 
\cite{ihle_03_srd_b}   
\begin{equation}\label{GK_sd} 
D= \frac{k_{B}T\tau}{2}\sum_{n=0}^{\infty}{^\prime} H (n\tau),  
\end{equation}
and  
\begin{equation}
H(n\tau) = \langle {\bf v}_{i}(0)\cdot {\bf v}_{i}(n\tau) \rangle  
/(k_{B}T)\;.
\end{equation}
Since the self-diffusion coefficient is a single-particle property, there is 
no sum over $i$ in Eq. (\ref{GK_sd}). In the molecular chaos approximation, 
\be\label{H_MC}   
H(n\tau) \equiv H_c(n\tau) = 2 \left[ \cos\left(\alpha\right) +\left( 1- \cos\left(\alpha\right) \right)
\left(1-e^{-M}\right)/M \right]^n  \;\;, 
\ee
Inserting Eq. (\ref{H_MC}) into Eq. (\ref{GK_sd}) and summing the resulting 
geometric series, one can obtain the expression for the self-diffusion 
coefficient in two dimensions given in Table \ref{tab_transport}. 

\begin{figure}
\begin{center}
\vspace{3cm}
\includegraphics[width=5in,angle=0]{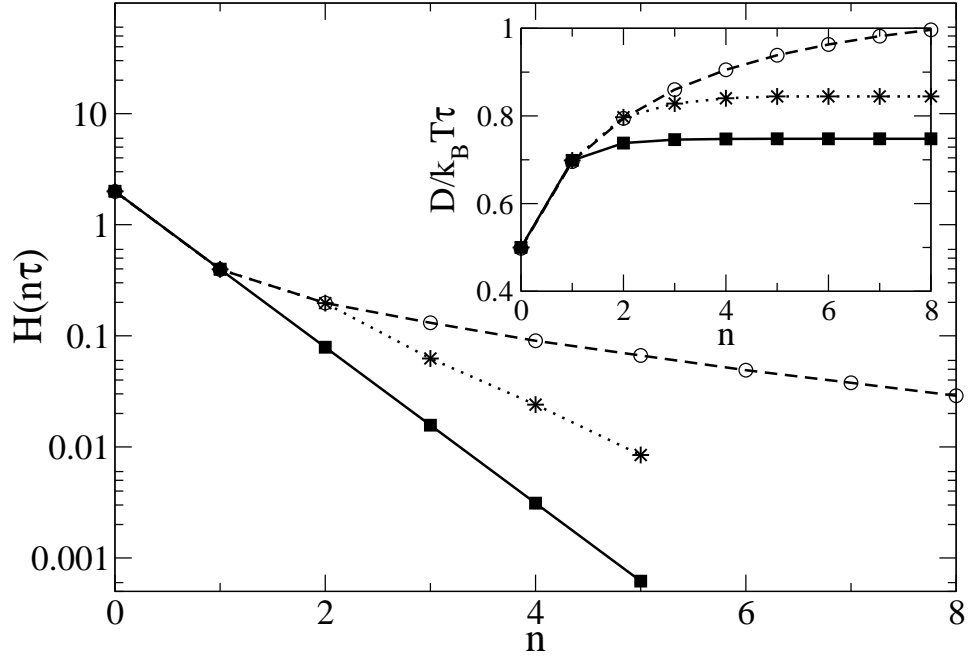}
\caption{Normalized velocity auto-correlation function $H(n\tau)$ as a 
function of time step $n$. Measurement values are shown by open circles 
($\smallwhitecircle$), the geometric series by filled squares 
($\smallblacksquare$), and the sum of the geometric series and the 
correlation contributions (Eq. \ref{Hsum}, using 
the numerically determined $p_2$) are shown by asterisks ($\smallstar$). 
The inset shows the normalized diffusion coefficient, $D/k_BT\tau$, as a 
function of the time step $n$.
The asterisks in the inset are a plot of Eq. (\ref{Dsum}) using Eq. (\ref{Hsum}).
Parameters: $\lambda/a=0.1$, $\alpha=90^{\circ}$, $L/a=64$, $M=5$ and 
$\tau=1.0$.}
\label{fig_correlationH}
\end{center}
\end{figure}

The self-diffusion coefficient is unique in that there is no ``collisional'' 
contribution; as a result, correlation corrections are much more important 
at small mean free path and can lead to large corrections to the 
results obtained using the molecular chaos approximation. 
Corrections to this result occur when two or more particles occupy the 
same collision cell at different time steps. Figure \ref{fig_correlationH} 
contains a comparison of the molecular chaos approximation for the 
velocity auto-correlation function, $H(n\tau)$, ($\smallblacksquare$) 
with simulation results ($\smallwhitecircle$). 

\begin{figure}
\begin{center}
\vspace{3.5cm}
\includegraphics[width=1in,angle=0]{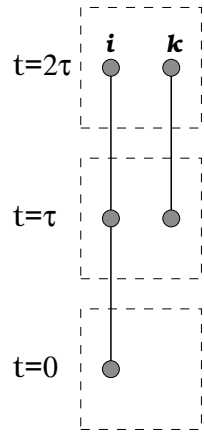}
\caption{Schematic diagram showing the configuration contributing  
to correlations at $t=2\tau$ in the calculation of self-diffusion coefficient. 
Particles $i$ and $k$ are in the same shifted collision cell at both $t=\tau$ 
and $t=2\tau$. There are $M-1$ such contributions with probability $p_2$.}
\label{fig_correlationdiff}
\end{center}
\end{figure}

The first of these correlation corrections, $\delta H(2\tau)$, occurs 
at $t=2\tau$. The contributing configuration, in which two particles, 
$i$ and $k$, are in the same (shifted) cell at both $t=\tau$ and $t=2\tau$, 
is shown in Figure \ref{fig_correlationdiff}; the probability for this to 
occur is again $p_2$. The contribution of this configuration  
to the velocity auto-correlation function is 
\begin{equation}\label{h2} 
\delta H(2\tau)  = 2 p_2  (M-1)\left\langle v_{ix}(0) 
\left[ \frac{1-c}{M}v_{kx}(\tau)-\frac{s}{M}v_{ky}(\tau)\right]\right\rangle
/(k_{B}T)\;,
\end{equation} 
where the factor 2 arises since both $x$ and $y$ components contribute; the 
factor $M-1$ accounts for the fact that $k\ne i$ can correspond to any 
of the $M-1$ particles. Following the procedure outlined in the discussion 
of correlation effects to the viscosity, it is straightforward to evaluate the 
averages in Eq. (\ref{h2}). The final result is 
\begin{equation}   
\delta H(2\tau)=2 p_2  (M-1)\left( \frac{1-c}{M} \right)^2 = 
\frac{8(M-1)}{9M^2}\left[1-\cos(\alpha)\right]^2\;. \label{deltaH}
\end{equation} 
This is the only correlation correction at $t=2\tau$. At longer times
there are similar higher order correlation effects arising, for example, 
when two particles are in the same shifted cell for three time steps, 
etc. It is straightforward but tedious to calculate these contributions.  
The probability $p_2$ in Eq. (\ref{deltaH}) is determined analytically in 
the limit $\lambda/a\to0$ in Appendix A, where it is shown that $p_2=(2/3)^d$ 
in $d$ dimensions. For finite $\lambda/a$, it can be measured in simulations. 
It should be noted that $p_2$ is related to the quantity $\zeta_1$ of 
Ref. ~\cite{ripoll_05_drf}, which denotes the number of particles that are 
neighbors of a given particle for two consecutive time steps; more 
precisely, $p_2=(\zeta_1 -1)/(M-1)$. 

Figure \ref{fig_H} is a plot of simulation results for $\delta H(2\tau)$ 
as a function of $\lambda/a$ for three different values of the collision 
angle $\alpha$. The asterisks ($\smallstar$) are result of Eq. (\ref{deltaH}) using 
the $\lambda/a\rightarrow 0$ prediction $p_2=4/9$; as can be seen, the agreement with 
simulation data is excellent. The inset in Figure \ref{fig_H} shows $p_2$ 
as a function of $\lambda/a$. The value for the probability $p_2$ in the 
limit $\lambda/a\to0$ is in excellent agreement with the result derived 
in Appendix A. 

\begin{figure}
\begin{center}
\vspace{3cm}
\includegraphics[width=2.8in,angle=0]{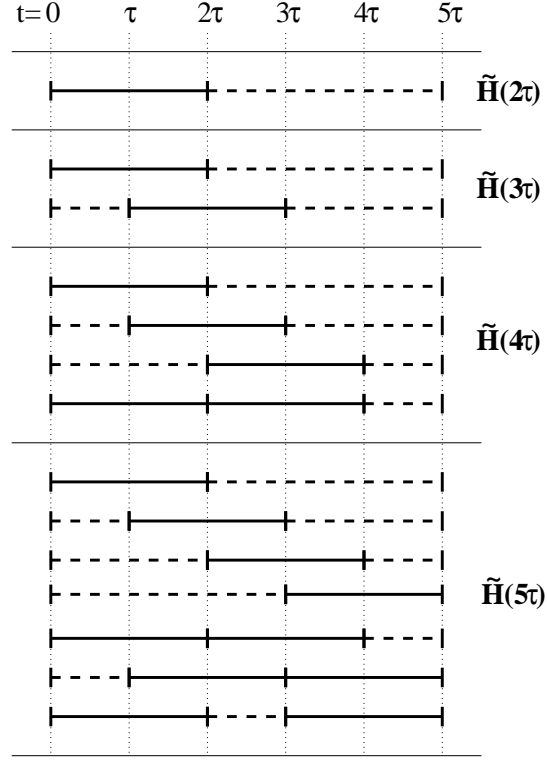}
\caption{Schematic diagram of contributions from $\delta H(2\tau)$.
Solid lines, all of which have a length of two time steps, show the time 
span during which particles stay in the same collision cell and are therefore 
correlated. Dashed lines show the time span in which particles are
uncorrelated, so that the molecular chaos assumption is valid. Each solid 
line will contribute a term $\delta H(2\tau)$ and dashed lines will 
contribute a factor $H_c(n\tau)$. 
$\delta \tilde{H}(n\tau)$ for $n=2,3,4,5$ show the first four contributions.  
Note that two consecutive solid lines, i.e. as shown in 
$\delta \tilde{H}(4\tau)$, means that particle $i$ is correlated 
twice for two consecutive time steps, but not with the same particle. }
\label{fig_corrsum}
\end{center}
\end{figure}

We have only considered correlation effects caused by two particles occupying 
the same collision cell for the two consecutive time steps. In fact, additional 
contributions arise any time two particles find themselves in the same 
collision cell for more than two time steps or after any number of time steps.
It is interesting to see just how important these latter contributions are 
by summing up all possible contributions from $\delta H(2\tau)$ and 
$H_c (\tau)$. Figure \ref{fig_corrsum} shows the contributions for the 
first five time steps in the series.   
In this approximation, the self-diffusion constant can be written as 
\begin{equation}
D\simeq\frac{k_BT\tau}{2} \sum_{n=0}^{\infty}{^\prime}\left[ H_c (n\tau)+\delta \tilde{H}(n\tau)\right] \label{Dsum}
\end{equation}
where 
\begin{equation}
 \delta \tilde{H}(n\tau)= \left\{ 
\begin{array}{cc}
0 ,&  n\leq1 \\
\delta H(2\tau),&  n=2 \\
\delta H(2\tau) H_c(\tau), & n=3 \\
(1/2)\delta H(2\tau)^2+(3/2)\delta H(2\tau)H_c(2\tau) , & n=4 \\
(3/4)\delta H(2\tau)^2 H_c(\tau)+2\delta H(2\tau)H_c(3\tau)  , & n=5
\end{array}\right. \label{Hsum}
\end{equation}
are the contributions shown in Figure \ref{fig_corrsum}. The solid lines 
in the figure represent the factor $\delta H(2\tau)$, and the dashed lines 
represent the factor $H_c(\tau)$ from time steps where particles are not 
correlated and the molecular chaos approximation is valid. The resulting 
contribution of Eq. (\ref{Hsum}) is shown in Figure \ref{fig_correlationH} 
by asterisks ($\smallstar$). The agreement with simulations at $t=2\tau$ 
is perfect, as expected, and the prediction for larger time intervals is 
improved. It would be extremely interesting to extend the analysis to 
consider the contributions of correlation affects at larger time intervals 
that lead to long time tails in the velocity auto-correlation function~\cite{dorfman_70_vcf}. 


\section{Conclusion}

This paper contains the first detailed analysis of equilibrium dynamic 
correlation functions using the SRD 
algorithm. The dynamic structure factor, vorticity and entropy density 
correlation functions were measured and used to provide unbiased estimates 
for the viscosity, thermal diffusivity, and bulk viscosity. The results are in 
good agreement with earlier numerical and theoretical results, and provide the 
first direct verification that the bulk viscosity is zero for this 
algorithm. 

Table \ref{tab_transport} contains a complete summary of analytical results 
for the transport coefficients of this model, and the results of this paper 
verify that we have an excellent understanding of the SRD algorithm at the 
kinetic level and that the analytic expressions for the transport coefficients 
do indeed provide a very accurate description of the SRD fluid. Furthermore, 
the analysis of the dynamical structure factor and the dynamic entropy density 
correlation function verify directly that the algorithm satisfies the 
fluctuation-dissipation theorem. 
Verification studies of this type will be important for generalizations 
of the algorithm which model excluded volume effects through the use of 
biased multi-particle collision rules which depend on the local velocities 
and densities. 

Finally, corrections to the self-diffusion coefficient and shear viscosity
arising from the breakdown of the molecular chaos approximation at small 
mean free paths were analyzed. In addition to deriving the form of 
the leading correlation corrections to these transport coefficients, 
the probabilities that two and three particles remain collision partners 
for consecutive time steps are derived analytically in the limit of small 
mean free path. Extensions of this approach could be used to study the 
development of long time tails in the velocity and stress auto-correlation 
functions.  

\begin{acknowledgments}
Support from the 
National Science Foundation under grant No. DMR-0513393 and  
ND EPSCoR through NSF grant EPS-0132289, is greatfully acknowledged. 
\end{acknowledgments}

\appendix

\section{Calculation of $p_2$}

Random grid shifts in $x$- and $y$-directions are statistically independent. We 
therefore first calculate the contribution from shifts in the $x$-direction; 
the final probability for general shifts in two-dimensions is then obtained 
by squaring this one-dimensional result. The following calculations are 
done in the limit $\lambda/a\to0$, so that the particles do not move between 
time steps.  

There are three different ways for two particles to be in the same 
shifted cell at consecutive time steps (see Figure \ref{fig_shift}). 
If the shifted cell index at 
$t=\tau$, $\xi_s(\tau)$ is zero, then $\xi_s(2\tau)$ can be either $-a,0$ 
or $a$. Because these are mutually exclusive events, the probability of each 
scenario has to be calculated separately and then summed. 

\subsection{\it $\xi_s(\tau)=0$, $\xi_s(2\tau)=0$ } \label{secp2A}

The situation is shown pictorially in Figure \ref{fig_shift}(a). The 
probability $p_{2}^{A}$ that two particles are in cell $\xi_s=0$ for two 
consecutive time steps can be written as
\begin{eqnarray}
p_2^A = \frac{1}{a^4} \idelta{2} \idelta{1} \ix{1} \ix{2}\!\!\!\!\!\!\!\! && \T\left(x_1 - \d_2 \right) \left[ 1-
\T\left(x_1-a-\d_2 \right) \right] \nonumber \\
&\times & \T\left(x_2 - \d_2 \right) \left[ 1-\T\left(x_2-a-\d_2 \right) 
\right] \;, \label{pA_1}
\end{eqnarray}
where $\delta_1$ and $\delta_2$ are the shifts at times $\tau$ and $2\tau$, 
respectively. Making the substitutions $X_1\equiv x_1-\d_1$ and 
$X_2\equiv x_2-\d_1$, Eq. (\ref{pA_1}) becomes
\begin{eqnarray}
p_2^A = \frac{1}{a^4} \idelta{2} \idelta{1} \iX{1} \iX{2}\!\!\!\!\!\!\!\! && 
\T\left(X_1 + \d_1 - \d_2 \right) \left[ 1-\T\left(X_1-a+\d_1 -\d_2 \right) 
\right] \nonumber \\
&\times & \T\left(X_2 +\d_1 - \d_2 \right) 
\left[ 1-\T\left(X_2-a+\d_1 -\d_2 \right) \right].  
\end{eqnarray}
To simplify further, 
introduce $\rho_1\equiv \d_1 -\d_2$ and integrate $X_1$ and $X_2$ over the 
portion of the square where the integrand is non-zero. This yields
\begin{equation}
p_2^A=\frac{1}{a^4}\idelta{2}\idelta{1} \left( a- |\rho_1|\right)^2 \;. 
\label{pA_2}
\end{equation}
\noindent
Using, finally, 
\begin{eqnarray}
\idelta{2} \idelta{1}  &=& a^{2}, \label{pA_3} \\
\idelta{2} \idelta{1} |\rho_1| &=& \frac{a^3}{3} \label{pA_4}\\
\idelta{2} \idelta{1} \rho_1^2 &=& \frac{a^4}{6}, \label{pA_5} 
\end{eqnarray}
Eq. (\ref{pA_2}) gives
\begin{equation}
p_2^A=1-\frac{2}{3}+\frac{1}{6}=\frac{1}{2} \;. \label{p2A}
\end{equation} 

\subsection{\it $\xi_s(\tau)=0$, $\xi_s(2\tau)=-a$ }\label{secp2B}

This situation is illustrated in Figure \ref{fig_shift}b. The probability 
$p_2^B$ that two particles are in cell $\xi_s=0$ at $t=\tau$ and $\xi_s=-a$ 
at $t=2\tau$ is
\begin{eqnarray}
p_2^B = \frac{1}{a^4} \idelta{2} \idelta{1} \ix{1} 
\ix{2}\!\!\!\!\!\!\!\! && \T\left(x_1 +a - \d_2 \right) \left[ 1-
\T\left(x_1-\d_2 \right) \right] \nonumber \\
&\times & \T\left(x_2 +a - \d_2 \right) \left[ 1-\T\left(x_2-\d_2 \right) \right] \;\;. \label{pB_1}
\end{eqnarray}
Making the substitutions $X_1\equiv x_1-\d_1$ and $X_2\equiv x_2-\d_1$, 
Eq. (\ref{pB_1}) becomes
\begin{eqnarray}
p_2^B = \frac{1}{a^4} \idelta{2} \idelta{1} \iX{1} \iX{2}\!\!\!\!\!\!\!\! 
&& \T\left(X_1 + a+ \d_1 - \d_2 \right) 
\left[ 1-\T\left(X_1+\d_1 -\d_2 \right) \right] \nonumber \\
&\times & \T\left(X_2 +a+\d_1 - \d_2 \right) \left[ 1-\T\left(X_2+\d_1 -\d_2 
\right) \right] \;.
\end{eqnarray}
As in the previous subsection, introducing $\rho_1\equiv \d_1 -\d_2$ and 
integrating $X_1$ and $X_2$ over the portion of the square where the integrand 
is non-zero yields 
\begin{equation}
p_2^B=\frac{1}{a^4}\idelta{2}\idelta{1} |\rho_1|^2 \T\left(-\rho_1\right)= 
\frac{1}{12}\;. \label{p2B}
\end{equation}

\subsection{\it $\xi_s(\tau)=0$, $\xi_s(2\tau)=a$ } \label{secp2C}

Referring to Figure \ref{fig_shift}c.
the probability $p_2^C$ that two particles are in 
cell $\xi_s=0$ at $t=\tau$ and $\xi_s=a$ at $t=2\tau$ is
\begin{eqnarray}
p_2^C = \frac{1}{a^4} \idelta{2} \idelta{1} \ix{1} \ix{2}\!\!\!\!\!\!\!\! 
&& \T\left(x_1 - a + \d_2 \right) \left[ 1-
\T\left(x_1- 2a + \d_2 \right) \right] \nonumber \\
&\times & \T\left(x_2 -a + \d_2 \right) 
\left[ 1-\T\left(x_2 -2a + \d_2 \right) \right] \;. \label{pC_1}
\end{eqnarray}
Making the same change of variables as in the previous two cases, 
Eq. (\ref{pC_1}) becomes
\begin{eqnarray}
p_2^C = \frac{1}{a^4} \idelta{2} \idelta{1} \iX{1} \iX{2}\!\!\!\!\!\!\!\! && \T\left(X_1 - a+ \d_1 + \d_2 \right) \left[ 1-\T\left(X_1 -2a +\d_1 + \d_2 \right) \right] \nonumber \\
&\times & \T\left(X_2  -a+\d_1 + \d_2 \right) 
\left[ 1-\T\left(X_2 -2a +\d_1 +\d_2 \right) \right] \;.
\end{eqnarray}
Introducing now $\rho_2\equiv \d_1 +\d_2$ and performing the integrals 
over $X_1$ and $X_2$ yields 
\begin{equation}
p_2^C=\frac{1}{a^4}\idelta{2}\idelta{1} {\rho_2}^2 
\T\left(\rho_2\right)= \frac{1}{12}\;. \label{p2C}
\end{equation}
The final result in two dimensions is obtained by summing the results 
given in Eqs. (\ref{p2A}), (\ref{p2B}) and (\ref{p2C}), and squaring, so 
that  
\begin{equation}
p_2=(p_2^A+p_2^B+p_2^C)^2=\frac{4}{9}\;.\label{p2}
\end{equation}

\section{Calculation of $p_3$}

\begin{figure}
\begin{center}
\vspace{3cm}
\includegraphics[width=3in,angle=0]{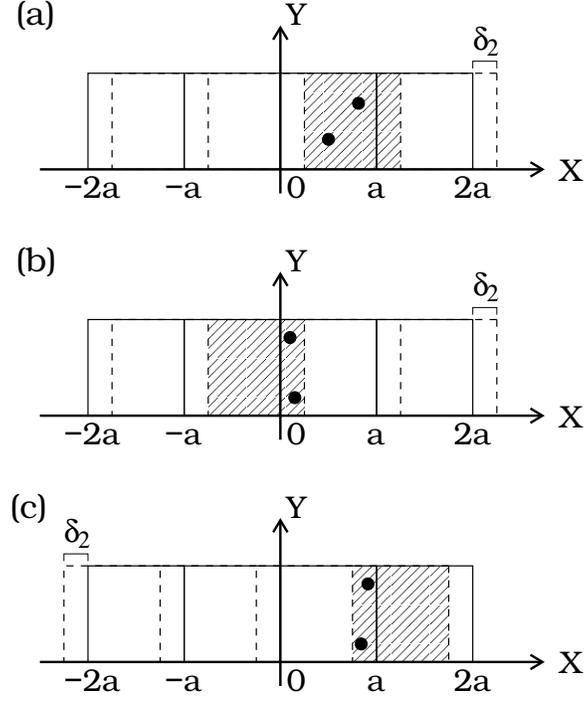}
\caption{Schematic diagram showing ways in which two particles can be in 
the same shifted cell at consecutive time steps. The boxes with solid 
and dashed borders represent the shifted grids at $t=\tau$ and $t=2\tau$, 
respectively. $\delta_2$ is the shift at $t=2\tau$. The coordinate system 
uses the shifted frame at $t=\tau$ as a reference. Two particles can be in 
a) the same shifted cell $\xi_s=0$ at both $t=\tau$ and $t=2\tau$, b) cells 
$\xi_s=0$ at $t=\tau$ and $\xi_s=-a$ at $t=2\tau$, or c) cells $\xi_s=0$ at 
$t=\tau$ and $\xi_s=a$ at $t=2\tau$. For simplicity, only shifts in 
$x$-direction are shown.
}
\label{fig_shift}
\end{center}
\end{figure}

The calculation of $p_3$ in the limit $\lambda/a\to0$ is similar to that of 
$p_2$ in the previous Appendix. There are three scenarios, as depicted in 
Figure \ref{fig_shift} (with three particles instead of two).  
$p_3^A$ is the probability that three particles are in the  
$\xi_s=0$ for two consecutive time steps: 
\begin{eqnarray}
p_3^A = \frac{1}{a^5} \idelta{2} \idelta{1} \ix{1} \ix{2} 
\ix{3}\!\!\!\!\!\!\!\! && \T\left(x_1 - \d_2 \right) \left[ 1-
\T\left(x_1-a-\d_2 \right) \right] \nonumber \\
&\times & \T\left(x_2 - \d_2 \right) \left[ 1-\T\left(x_2-a-\d_2 \right) \right] \nonumber \\
&\times & \T\left(x_3 - \d_2 \right) \left[ 1-\T\left(x_3-a-\d_2 \right) 
\right] \;.\label{p3A_1}
\end{eqnarray}
To evaluate this integral, make the same change of variables to $X_1$ and 
$X_2$ as in the previous Appendix and introduce $\rho_1=\delta_1-\delta_2$. 
Performing the $X$ integrals then gives 
\begin{equation}
p_3^A=\frac{1}{a^5}\idelta{2}\idelta{1} \left( a- |\rho_1|\right)^3 \;. 
\label{p3A_2}
\end{equation}
Using
\begin{equation}
\idelta{2} \idelta{1} |\rho_1|^3 = \frac{a^5}{10} \label{p3A_3}
\end{equation}
and Eqs. (\ref{pA_3}), (\ref{pA_4}), (\ref{pA_5}) and (\ref{p3A_3}), 
Eq. (\ref{p3A_2}) yields 
\begin{equation}
p_3^A=1-1+\frac{1}{2}-\frac{1}{10}=\frac{2}{5} \;. \label{pA_{3}}
\end{equation} 

The calculations of $p_3^B$ and $p_3^C$ are similar to those outlined in 
Sections \ref{secp2B} and \ref{secp2C}, and both are equal to $1/20$. 
Summing these results, 
\begin{equation}
p_{3}=\left(p_{3}^{A}+p_{3}^{B}+p_{3}^{C}\right)^{2}=\frac{1}{4} \label{p3}
\end{equation}
in two dimensions. 

\bibliography{TuzelLibrary}

\begin{thebibliography}{34}
\expandafter\ifx\csname natexlab\endcsname\relax\def\natexlab#1{#1}\fi
\expandafter\ifx\csname bibnamefont\endcsname\relax
  \def\bibnamefont#1{#1}\fi
\expandafter\ifx\csname bibfnamefont\endcsname\relax
  \def\bibfnamefont#1{#1}\fi
\expandafter\ifx\csname citenamefont\endcsname\relax
  \def\citenamefont#1{#1}\fi
\expandafter\ifx\csname url\endcsname\relax
  \def\url#1{\texttt{#1}}\fi
\expandafter\ifx\csname urlprefix\endcsname\relax\def\urlprefix{URL }\fi
\providecommand{\bibinfo}[2]{#2}
\providecommand{\eprint}[2][]{\url{#2}}

\bibitem[{\citenamefont{Malevanets and Kapral}(1999)}]{malevanets_99_mms}
\bibinfo{author}{\bibfnamefont{A.}~\bibnamefont{Malevanets}} \bibnamefont{and}
  \bibinfo{author}{\bibfnamefont{R.}~\bibnamefont{Kapral}},
  \bibinfo{journal}{J. Chem. Phys.} \textbf{\bibinfo{volume}{110}},
  \bibinfo{pages}{8605} (\bibinfo{year}{1999}).

\bibitem[{\citenamefont{Malevanets and Kapral}(2000)}]{malevanets_00_smd}
\bibinfo{author}{\bibfnamefont{A.}~\bibnamefont{Malevanets}} \bibnamefont{and}
  \bibinfo{author}{\bibfnamefont{R.}~\bibnamefont{Kapral}},
  \bibinfo{journal}{J. Chem. Phys.} \textbf{\bibinfo{volume}{112}},
  \bibinfo{pages}{7260} (\bibinfo{year}{2000}).

\bibitem[{\citenamefont{Bird}(1994)}]{bird_94_mgd}
\bibinfo{author}{\bibfnamefont{G.~A.} \bibnamefont{Bird}},
  \emph{\bibinfo{title}{Molecular Gas Dynamics and the Direct Simulation of Gas
  Flows}} (\bibinfo{publisher}{Oxford University Press, USA},
  \bibinfo{year}{1994}).

\bibitem[{\citenamefont{Ihle and Kroll}(2001)}]{ihle_01_srd}
\bibinfo{author}{\bibfnamefont{T.}~\bibnamefont{Ihle}} \bibnamefont{and}
  \bibinfo{author}{\bibfnamefont{D.~M.} \bibnamefont{Kroll}},
  \bibinfo{journal}{Phys. Rev. E} \textbf{\bibinfo{volume}{63}},
  \bibinfo{pages}{020201} (\bibinfo{year}{2001}).

\bibitem[{\citenamefont{Pooley and Yeomans}(2005)}]{pooley_05_ktd}
\bibinfo{author}{\bibfnamefont{C.~M.} \bibnamefont{Pooley}} \bibnamefont{and}
  \bibinfo{author}{\bibfnamefont{J.~M.} \bibnamefont{Yeomans}},
  \bibinfo{journal}{J. Phys. Chem. B} \textbf{\bibinfo{volume}{109}},
  \bibinfo{pages}{6505} (\bibinfo{year}{2005}).

\bibitem[{\citenamefont{Ryder}(2005)}]{ryder_05_thesis}
\bibinfo{author}{\bibfnamefont{J.~F.} \bibnamefont{Ryder}}, Ph.D. thesis,
  \bibinfo{school}{University of Oxford} (\bibinfo{year}{2005}).

\bibitem[{\citenamefont{Kikuchi et~al.}(2002)\citenamefont{Kikuchi, Gent, and
  Yeomans}}]{kikuchi_02_pcp}
\bibinfo{author}{\bibfnamefont{N.}~\bibnamefont{Kikuchi}},
  \bibinfo{author}{\bibfnamefont{A.}~\bibnamefont{Gent}}, \bibnamefont{and}
  \bibinfo{author}{\bibfnamefont{J.~M.} \bibnamefont{Yeomans}},
  \bibinfo{journal}{Eur. Phys. J. E} \textbf{\bibinfo{volume}{9}},
  \bibinfo{pages}{63} (\bibinfo{year}{2002}).

\bibitem[{\citenamefont{Winkler et~al.}(2004)\citenamefont{Winkler,
  Mussawisade, Ripoll, and Gompper}}]{winkler_04_rcp}
\bibinfo{author}{\bibfnamefont{R.~G.} \bibnamefont{Winkler}},
  \bibinfo{author}{\bibfnamefont{K.}~\bibnamefont{Mussawisade}},
  \bibinfo{author}{\bibfnamefont{M.}~\bibnamefont{Ripoll}}, \bibnamefont{and}
  \bibinfo{author}{\bibfnamefont{G.}~\bibnamefont{Gompper}},
  \bibinfo{journal}{J. Phys.: Condens. Matter} \textbf{\bibinfo{volume}{16}},
  \bibinfo{pages}{3941} (\bibinfo{year}{2004}).

\bibitem[{\citenamefont{Webster and Yeomans}(2005)}]{webster_05_mtp}
\bibinfo{author}{\bibfnamefont{M.~A.} \bibnamefont{Webster}} \bibnamefont{and}
  \bibinfo{author}{\bibfnamefont{J.~M.} \bibnamefont{Yeomans}},
  \bibinfo{journal}{J. Chem. Phys.} \textbf{\bibinfo{volume}{122}},
  \bibinfo{pages}{164903} (\bibinfo{year}{2005}).

\bibitem[{\citenamefont{Falck et~al.}(2004)\citenamefont{Falck, Lahtinen,
  Vattulainen, and Ala-Nissila}}]{falck_04_ihm}
\bibinfo{author}{\bibfnamefont{E.}~\bibnamefont{Falck}},
  \bibinfo{author}{\bibfnamefont{J.~M.} \bibnamefont{Lahtinen}},
  \bibinfo{author}{\bibfnamefont{I.}~\bibnamefont{Vattulainen}},
  \bibnamefont{and}
  \bibinfo{author}{\bibfnamefont{T.}~\bibnamefont{Ala-Nissila}},
  \bibinfo{journal}{Eur. Phys. J. E} \textbf{\bibinfo{volume}{13}},
  \bibinfo{pages}{267} (\bibinfo{year}{2004}).

\bibitem[{\citenamefont{Inoue et~al.}(2002)\citenamefont{Inoue, Chen, and
  Ohashi}}]{inoue_02_dsm}
\bibinfo{author}{\bibfnamefont{Y.}~\bibnamefont{Inoue}},
  \bibinfo{author}{\bibfnamefont{Y.}~\bibnamefont{Chen}}, \bibnamefont{and}
  \bibinfo{author}{\bibfnamefont{H.}~\bibnamefont{Ohashi}},
  \bibinfo{journal}{J. Stat. Phys.} \textbf{\bibinfo{volume}{107}},
  \bibinfo{pages}{85} (\bibinfo{year}{2002}).

\bibitem[{\citenamefont{Padding and Louis}(2004)}]{padding_04_hbf}
\bibinfo{author}{\bibfnamefont{J.~T.} \bibnamefont{Padding}} \bibnamefont{and}
  \bibinfo{author}{\bibfnamefont{A.~A.} \bibnamefont{Louis}},
  \bibinfo{journal}{Phys. Rev. Lett.} \textbf{\bibinfo{volume}{93}},
  \bibinfo{pages}{220601} (\bibinfo{year}{2004}).

\bibitem[{\citenamefont{Hecht et~al.}(2005)\citenamefont{Hecht, Harting, Ihle,
  and Herrmann}}]{hecht_05_scc}
\bibinfo{author}{\bibfnamefont{M.}~\bibnamefont{Hecht}},
  \bibinfo{author}{\bibfnamefont{J.}~\bibnamefont{Harting}},
  \bibinfo{author}{\bibfnamefont{T.}~\bibnamefont{Ihle}}, \bibnamefont{and}
  \bibinfo{author}{\bibfnamefont{H.~J.} \bibnamefont{Herrmann}},
  \bibinfo{journal}{Phys. Rev. E} \textbf{\bibinfo{volume}{72}},
  \bibinfo{pages}{011408} (\bibinfo{year}{2005}).

\bibitem[{\citenamefont{Padding and Louis}()}]{padding_06_hib}
\bibinfo{author}{\bibfnamefont{J.~T.} \bibnamefont{Padding}} \bibnamefont{and}
  \bibinfo{author}{\bibfnamefont{A.~A.} \bibnamefont{Louis}},
  \bibinfo{note}{cond-mat/0603391}.

\bibitem[{\citenamefont{Noguchi and Gompper}(2004)}]{noguchi_04_fvw}
\bibinfo{author}{\bibfnamefont{H.}~\bibnamefont{Noguchi}} \bibnamefont{and}
  \bibinfo{author}{\bibfnamefont{G.}~\bibnamefont{Gompper}},
  \bibinfo{journal}{Phys. Rev. Lett.} \textbf{\bibinfo{volume}{93}},
  \bibinfo{pages}{258102} (\bibinfo{year}{2004}).

\bibitem[{\citenamefont{Noguchi and Gompper}(2005)}]{noguchi_05_dfv}
\bibinfo{author}{\bibfnamefont{H.}~\bibnamefont{Noguchi}} \bibnamefont{and}
  \bibinfo{author}{\bibfnamefont{G.}~\bibnamefont{Gompper}},
  \bibinfo{journal}{Phys. Rev. E} \textbf{\bibinfo{volume}{72}},
  \bibinfo{pages}{011901} (\bibinfo{year}{2005}).

\bibitem[{\citenamefont{Ihle and Kroll}(2003{\natexlab{a}})}]{ihle_03_srd_a}
\bibinfo{author}{\bibfnamefont{T.}~\bibnamefont{Ihle}} \bibnamefont{and}
  \bibinfo{author}{\bibfnamefont{D.~M.} \bibnamefont{Kroll}},
  \bibinfo{journal}{Phys. Rev. E} \textbf{\bibinfo{volume}{67}},
  \bibinfo{pages}{066705} (\bibinfo{year}{2003}{\natexlab{a}}).

\bibitem[{\citenamefont{Kikuchi et~al.}(2003)\citenamefont{Kikuchi, Pooley,
  Ryder, and Yeomans}}]{kikuchi_03_tcm}
\bibinfo{author}{\bibfnamefont{N.}~\bibnamefont{Kikuchi}},
  \bibinfo{author}{\bibfnamefont{C.~M.} \bibnamefont{Pooley}},
  \bibinfo{author}{\bibfnamefont{J.~F.} \bibnamefont{Ryder}}, \bibnamefont{and}
  \bibinfo{author}{\bibfnamefont{J.~M.} \bibnamefont{Yeomans}},
  \bibinfo{journal}{J. Chem. Phys.} \textbf{\bibinfo{volume}{119}},
  \bibinfo{pages}{6388} (\bibinfo{year}{2003}).

\bibitem[{\citenamefont{Ihle and Kroll}(2003{\natexlab{b}})}]{ihle_03_srd_b}
\bibinfo{author}{\bibfnamefont{T.}~\bibnamefont{Ihle}} \bibnamefont{and}
  \bibinfo{author}{\bibfnamefont{D.~M.} \bibnamefont{Kroll}},
  \bibinfo{journal}{Phys. Rev. E} \textbf{\bibinfo{volume}{67}},
  \bibinfo{eid}{066706} (\bibinfo{year}{2003}{\natexlab{b}}).

\bibitem[{\citenamefont{T{\"u}zel et~al.}(2003)\citenamefont{T{\"u}zel,
  Strauss, Ihle, and Kroll}}]{tuzel_03_tcs}
\bibinfo{author}{\bibfnamefont{E.}~\bibnamefont{T{\"u}zel}},
  \bibinfo{author}{\bibfnamefont{M.}~\bibnamefont{Strauss}},
  \bibinfo{author}{\bibfnamefont{T.}~\bibnamefont{Ihle}}, \bibnamefont{and}
  \bibinfo{author}{\bibfnamefont{D.~M.} \bibnamefont{Kroll}},
  \bibinfo{journal}{Phys. Rev. E} \textbf{\bibinfo{volume}{68}},
  \bibinfo{pages}{036701} (\bibinfo{year}{2003}).

\bibitem[{\citenamefont{Ihle et~al.}(2004)\citenamefont{Ihle, T{\"u}zel, and
  Kroll}}]{ihle_04_rgr}
\bibinfo{author}{\bibfnamefont{T.}~\bibnamefont{Ihle}},
  \bibinfo{author}{\bibfnamefont{E.}~\bibnamefont{T{\"u}zel}},
  \bibnamefont{and} \bibinfo{author}{\bibfnamefont{D.~M.} \bibnamefont{Kroll}},
  \bibinfo{journal}{Phys. Rev. E} \textbf{\bibinfo{volume}{70}},
  \bibinfo{pages}{035701} (\bibinfo{year}{2004}).

\bibitem[{\citenamefont{Ihle et~al.}(2005)\citenamefont{Ihle, T{\"u}zel, and
  Kroll}}]{ihle_05_ect}
\bibinfo{author}{\bibfnamefont{T.}~\bibnamefont{Ihle}},
  \bibinfo{author}{\bibfnamefont{E.}~\bibnamefont{T{\"u}zel}},
  \bibnamefont{and} \bibinfo{author}{\bibfnamefont{D.~M.} \bibnamefont{Kroll}},
  \bibinfo{journal}{Phys. Rev. E} \textbf{\bibinfo{volume}{72}},
  \bibinfo{pages}{046707} (\bibinfo{year}{2005}).

\bibitem[{\citenamefont{Allahyarov and Gompper}(2002)}]{allahyarov_02_mss}
\bibinfo{author}{\bibfnamefont{E.}~\bibnamefont{Allahyarov}} \bibnamefont{and}
  \bibinfo{author}{\bibfnamefont{G.}~\bibnamefont{Gompper}},
  \bibinfo{journal}{Phys. Rev. E} \textbf{\bibinfo{volume}{66}},
  \bibinfo{pages}{036702} (\bibinfo{year}{2002}).

\bibitem[{\citenamefont{Inoue et~al.}(2003)\citenamefont{Inoue, Chen, and
  Ohashi}}]{inoue_03_dcs}
\bibinfo{author}{\bibfnamefont{Y.}~\bibnamefont{Inoue}},
  \bibinfo{author}{\bibfnamefont{Y.}~\bibnamefont{Chen}}, \bibnamefont{and}
  \bibinfo{author}{\bibfnamefont{H.}~\bibnamefont{Ohashi}},
  \bibinfo{journal}{Comp. Phys. Commun.} \textbf{\bibinfo{volume}{153}},
  \bibinfo{pages}{66} (\bibinfo{year}{2003}).

\bibitem[{\citenamefont{Ripoll et~al.}(2004)\citenamefont{Ripoll, Mussawisade,
  Winkler, and Gompper}}]{ripoll_04_lhc}
\bibinfo{author}{\bibfnamefont{M.}~\bibnamefont{Ripoll}},
  \bibinfo{author}{\bibfnamefont{K.}~\bibnamefont{Mussawisade}},
  \bibinfo{author}{\bibfnamefont{R.~G.} \bibnamefont{Winkler}},
  \bibnamefont{and} \bibinfo{author}{\bibfnamefont{G.}~\bibnamefont{Gompper}},
  \bibinfo{journal}{Europhys. Lett.} \textbf{\bibinfo{volume}{68}},
  \bibinfo{pages}{106} (\bibinfo{year}{2004}).

\bibitem[{\citenamefont{Ripoll et~al.}(2005)\citenamefont{Ripoll, Mussawisade,
  Winkler, and Gompper}}]{ripoll_05_drf}
\bibinfo{author}{\bibfnamefont{M.}~\bibnamefont{Ripoll}},
  \bibinfo{author}{\bibfnamefont{K.}~\bibnamefont{Mussawisade}},
  \bibinfo{author}{\bibfnamefont{R.~G.} \bibnamefont{Winkler}},
  \bibnamefont{and} \bibinfo{author}{\bibfnamefont{G.}~\bibnamefont{Gompper}},
  \bibinfo{journal}{Phys. Rev. E} \textbf{\bibinfo{volume}{72}},
  \bibinfo{pages}{016701} (\bibinfo{year}{2005}).

\bibitem[{\citenamefont{Forster}(1975)}]{forster_75_hfb}
\bibinfo{author}{\bibfnamefont{D.}~\bibnamefont{Forster}},
  \emph{\bibinfo{title}{Hydrodynamic Fluctuations, Broken Symmetry, and
  Correlation Functions}} (\bibinfo{publisher}{W.A. Benjamin},
  \bibinfo{address}{Reading}, \bibinfo{year}{1975}).

\bibitem[{\citenamefont{Ihle et~al.}(2006)\citenamefont{Ihle, T{\"u}zel, and
  Kroll}}]{ihle_06_cpa}
\bibinfo{author}{\bibfnamefont{T.}~\bibnamefont{Ihle}},
  \bibinfo{author}{\bibfnamefont{E.}~\bibnamefont{T{\"u}zel}},
  \bibnamefont{and} \bibinfo{author}{\bibfnamefont{D.~M.} \bibnamefont{Kroll}},
  \bibinfo{journal}{Europhys. Lett.} \textbf{\bibinfo{volume}{73}},
  \bibinfo{pages}{664} (\bibinfo{year}{2006}).

\bibitem[{\citenamefont{T{\"u}zel et~al.}()\citenamefont{T{\"u}zel, Ihle, and
  Kroll}}]{tuzel_06_ctc}
\bibinfo{author}{\bibfnamefont{E.}~\bibnamefont{T{\"u}zel}},
  \bibinfo{author}{\bibfnamefont{T.}~\bibnamefont{Ihle}}, \bibnamefont{and}
  \bibinfo{author}{\bibfnamefont{D.~M.} \bibnamefont{Kroll}},
  \bibinfo{note}{cond-mat/0511312, to be published in Math. Comp. Simul.}

\bibitem[{\citenamefont{Boon and Yip}(1992)}]{boon_92_mh}
\bibinfo{author}{\bibfnamefont{J.~P.} \bibnamefont{Boon}} \bibnamefont{and}
  \bibinfo{author}{\bibfnamefont{S.}~\bibnamefont{Yip}},
  \emph{\bibinfo{title}{Molecular Hydrodynamics}} (\bibinfo{publisher}{Dover
  Publications, New York}, \bibinfo{year}{1992}).

\bibitem[{\citenamefont{Berne and Pecora}(1976)}]{berne_00_dls}
\bibinfo{author}{\bibfnamefont{B.~J.} \bibnamefont{Berne}} \bibnamefont{and}
  \bibinfo{author}{\bibfnamefont{R.}~\bibnamefont{Pecora}},
  \emph{\bibinfo{title}{Dynamic light scattering: with applications to
  chemistry, biology and physics}} (\bibinfo{publisher}{John Wiley \& Sons, New
  York}, \bibinfo{year}{1976}).

\bibitem[{\citenamefont{Tekmen}()}]{tekmen_private}
\bibinfo{author}{\bibfnamefont{M.}~\bibnamefont{Tekmen}},
  \emph{\bibinfo{title}{private communication}}.

\bibitem[{\citenamefont{Dorfman and Cohen}(1970)}]{dorfman_70_vcf}
\bibinfo{author}{\bibfnamefont{J.~R.} \bibnamefont{Dorfman}} \bibnamefont{and}
  \bibinfo{author}{\bibfnamefont{E.~G.~D.} \bibnamefont{Cohen}},
  \bibinfo{journal}{Phys. Rev. Lett.} \textbf{\bibinfo{volume}{25}},
  \bibinfo{pages}{1257} (\bibinfo{year}{1970}).

\bibitem[{\citenamefont{Pooley}(2003)}]{pooley_03_thesis}
\bibinfo{author}{\bibfnamefont{C.~M.} \bibnamefont{Pooley}}, Ph.D. thesis,
  \bibinfo{school}{University of Oxford} (\bibinfo{year}{2003}).

\end{thebibliography}

\newpage

\begin{table}
\caption{Theoretical expressions for the shear viscosity $\nu$, the thermal 
diffusivity $D_T$, and the self-diffusion coefficient $D$, in both two 
and three dimensions. $M$ denotes the average number of particles per cell, 
$\alpha$ is the collision angle,  $k_B$ is Boltzman's constant, $T$ is the 
temperature and $\tau$ is the time step. Except for self-diffusion constant, 
where there is no collisional contribution, both the kinetic and collisional 
contributions are listed. The expressions for shear viscosity and 
self-diffusion coefficient include fluctuation corrections for 
small $M$; however, for brevity the relations for thermal diffusivity are 
correct only up to $O(1/M)$ and $O(1/M^2)$ for the kinetic and collisional 
contributions, respectively. Both kinetic and collisional contributions to 
shear viscosity have been calculated using two complementary approaches, 
equilibrium Green-Kubo relations 
\cite{ihle_03_srd_a,ihle_03_srd_b,tuzel_03_tcs,ihle_04_rgr,ihle_05_ect} and a 
non-equilibrium approach~\cite{kikuchi_03_tcm,pooley_05_ktd}. Results from 
both approaches are in complete agreement. Similarly, the kinetic contribution 
to thermal diffusivity has also been calculated using these two approaches. 
The predictions of both approaches are identical in two dimensions and agree 
up to (and including) $O(1/M)$ in three dimensions; higher order contributions 
in $1/M$ were not considered in the Green-Kubo approach. The 
kinetic contribution to thermal diffusivity calculated using the non-equilibrium approach 
was taken from 
Ref. ~\cite{pooley_03_thesis}, since there appear to be several  
misprints in Eqs. (62), (63) and (64) in Ref. ~\cite{pooley_05_ktd}. 
The collisional contribution to thermal diffusivity has been calculated in 
both two and three dimensions in Refs. \cite{ihle_04_rgr,ihle_05_ect}. 
Because of space limitations, only the leading terms in $1/M$ are given here. 
For the complete expression, the reader is referred  
to ~\cite{ihle_05_ect}. To the best of our knowledge, the fluctuation 
corrections for self-diffusion coefficient are presented here for the first 
time.}
\def\arraystretch{2}
\begin{ruledtabular}
\begin{tabular}{|c|c|c|c|}
Transport coefficient &Dimension (d)~~& Kinetic$~~(\times k_BT\tau/2)$ & Collisional$~~(\times a^2/\tau)$ \\ \hline \hline
\multirow{2}{*}{Shear viscosity, $\nu$} & 2~ & $\frac{M}{(M-1+e^{-M})\sin^2(\alpha)}-1$ & \multirow{2}{*}{$\frac{1}{6 d M} \left({M-1+e^{-M}}\right)[1-\cos(\alpha)]$} \\ 
&3~ & $ \frac{5M}{(M-1+e^{-M})[2-\cos(\alpha)-\cos(2\alpha)]}-1 $ & \\ \hline
\multirow{2}{*}{Thermal diffusivity, $D_T$} & 2~ &
\multirow{2}{*}{$ \frac{d}{1-\cos(\alpha)} -1 
    + \frac{2d}{M} \left(\frac{7-d}{5} - \frac{1}{4} \,{\csc^2(\alpha/2)} \right)$}
 &  \multirow{2}{*}{$\frac{1}{3 (d+2)M }\left(1 - \frac{1}{M}\right) 
[1-\cos(\alpha)] $}    \\ 
& 3~ &  & \\ \hline
\multirow{2}{*}{Diffusion coefficient, $D$} & 2~ &\multirow{2}{*}{$\frac{dM}{(1-\cos(\alpha))(M-1+e^{-M})}  
 -1$} & \multirow{2}{*}{-} \\ 
& 3~ &  & \\
\end{tabular}
\label{tab_transport} 
\end{ruledtabular}
\end{table}

\newpage

\newpage

\newpage

\newpage

\newpage

\newpage

\newpage

\newpage

\newpage

\newpage

\newpage

\newpage

\newpage ~

\newpage
\begin{figure}
\begin{center}
\vspace{1.7cm}
\includegraphics[width=5in,angle=0]{deltaH_vs_lambda_insetp2.eps}
\caption{Correlation contributions at $t=2\tau$ as a function of 
$\lambda/a$ for the self-diffusion coefficient ($\delta H(2\tau)$). Results for  
collision angles $\alpha=60^{\circ}$ ($\smallwhitecircle$), $90^{\circ}$ 
($\smallwhitesquare$),  
and $120^{\circ}$ ($\smallwhitetriangleright$) are presented. The open and 
filled symbols represent data obtained for $\tau=1.0$ ($k_BT$ varied) and $k_BT=1.0$ ($\tau$ varied) respectively. The asterisks ($\smallstar$) are the 
theoretical predictions in the limit $\lambda/a\rightarrow 0$. The inset shows the numerically obtained probability $p_2$ as a function of mean free path for the different
parameters considered. Parameters: $L/a=64$, $M=5$.}
\label{fig_H}
\end{center}
\end{figure}

\newpage

\newpage

\end{document}